# Picosecond transfer from short-term to long-term memory in analog antiferromagnetic memory device

Miloslav Surýnek[1,4], Jan Zubáč[2,1,4], Kamil Olejník[2], Andrej Farkaš[2,1], Filip Křížek[2], Lukáš Nádvorník[1], Peter Kubaščík[1], František Trojánek[1], Richard P. Campion[3], Vít Novák[2], Tomáš Jungwirth[2,3], and Petr Němec[1,5*]

[1]Faculty of Mathematics and Physics, Charles University, Ke Karlovu 3, 121 16 Prague 2, Czech Republic

[2]Institute of Physics ASCR, v.v.i., Cukrovarnická 10, 162 53 Prague 6, Czech Republic

[3]School of Physics and Astronomy, University of Nottingham, Nottingham NG7 2RD, United Kingdom

[4]These authors contributed equally

[5*]Lead and corresponding author: petr.nemec@matfyz.cuni.cz

**Previous experiments in compensated magnets have demonstrated a potential for approaching the limit of fastest and least-dissipative operation of digital memory bits. However, the analog route has been virtually unexplored at (sub)ps time scales. In this paper, we report on experimental separation of heat-related and quench-switching-related resistance signal dynamics induced at room temperature by a single femtosecond-laser-pulse in memory devices made from a metallic antiferromagnetic CuMnAs. We show that the heat-related dynamics, on picosecond to hundreds of nanoseconds time scales, can be used as a short-term memory where information about input stimuli, represented by laser-pulses, is stored temporarily. When the quench-switching threshold is reached, information is transferred to the device's variable resistance, serving as a long-term memory, with time components of 10 ms and 10 s. The potentially appealing features of the heat-based memory for several distinct research fields, including bio-inspired analogue devices and heat-based logic, are also discussed.**

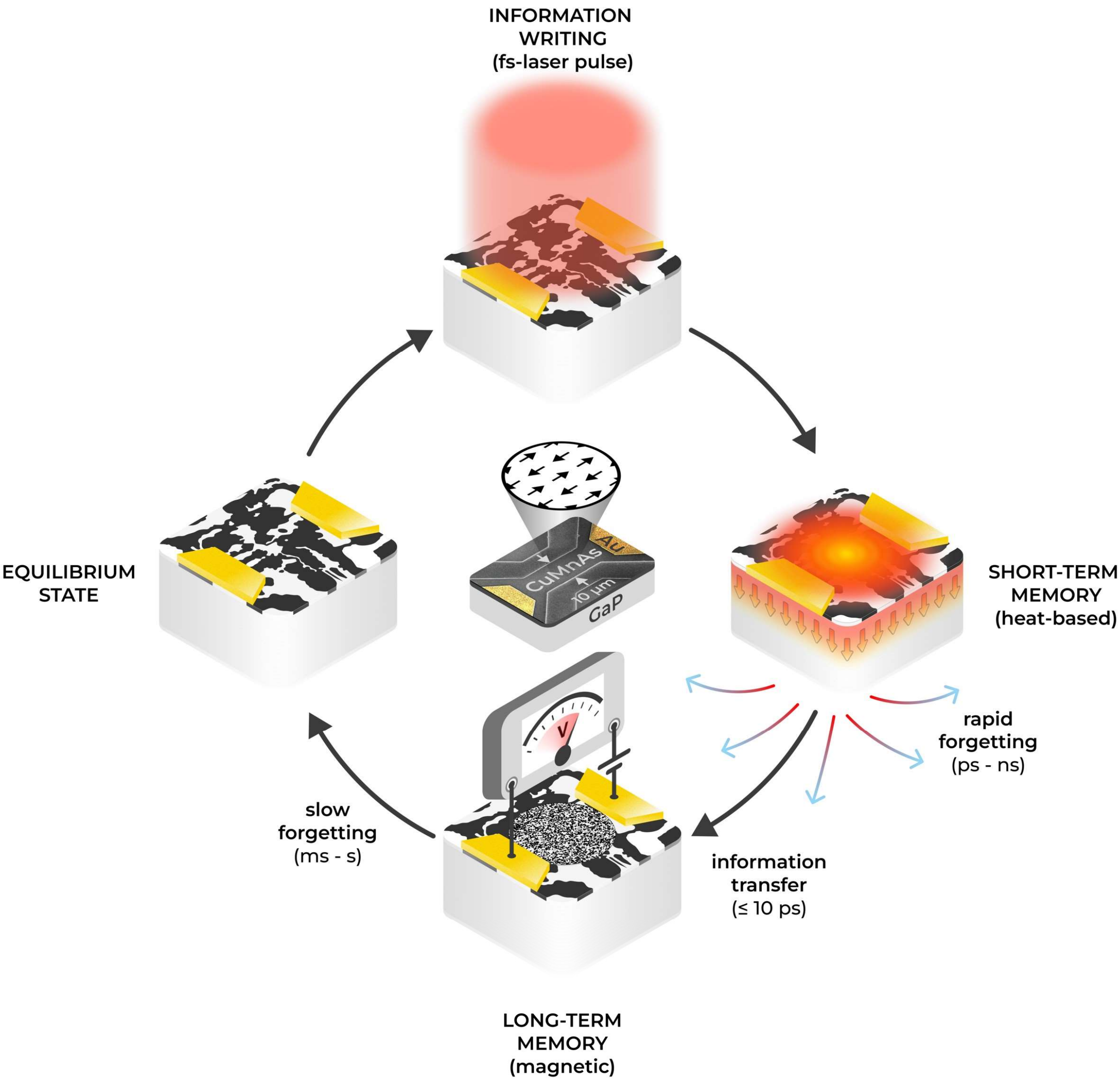

INFORMATION
WRITING
(fs-laser pulse)
EQUILIBRIUM
STATE
CuMnAs
Au
10 µm
GaP
SHORT-TERM
MEMORY
(heat-based)
rapid
forgetting
(ps - ns)
V
slow
forgetting
(ms - s)
information
transfer
(≤ 10 ps)
LONG-TERM
MEMORY
(magnetic)

## INTRODUCTION

The internal spin-dynamics in crystals with a compensated antiparallel ordering of magnetic moments is commonly in the ps (THz) range[1]. It makes these systems favorable for a demonstration of a memory technology approaching the thermodynamic Landauer ~ meV limit at room temperature, combined with the corresponding ~ ps temporal limit, of the least-dissipative and fastest operation of a digital bit[2,3]. A ps-scale Néel vector reorientation from a stable easy axis to a transiently excited distinct easy axis at energy of ~ meV per atom was demonstrated using a THz-field pulse in an insulating magnetically-compensated orthoferrite[4]. By employing a Néel spin-orbit torque excitation in metallic antiferromagnets, a method originally introduced in studies of electrical switching[5-7], it has been recently shown that a ps-scale switching between two stable Néel-vector orientations can be achieved by a THz field pulse at an energy of ~ 10 meV per atom[8]. These examples highlight that the ambition to approach the physical limit of the least-dissipative and fastest digital memory operation becomes a realistic, albeit still highly challenging objective of the research of ultra-fast magnetism in the compensated magnets.

In traditional digital computers based on the von Neumann architecture, physically separated processing and memory units are used. Consequently, the frequent data shuttling between them leads to considerable penalties on the energy efficiency and data bandwidth, termed the von Neumann bottleneck. The recent explosive growth of data-demanding applications related to Internet of Things and artificial intelligence calls for a change of the traditional computational schemes[9-12]. One such non-von Neumann computational approach is in-memory computing where certain computational tasks are performed in the memory itself[13-15]. Another extensively explored alternative is a bio-inspired neuromorphic computing using multi-state memory cells[16-18]. Moreover, many internet-connected devices intrinsically produce analog-valued data, such as pixel intensities from cameras, sound recordings from microphones, and time-series from sensors. This data type calls for an analog coding in emerging memory systems[19], which can also reduce latency by eliminating the need for multiple conversions between analog and digital data forms. From all these perspectives, metallic antiferromagnetic memory devices are very appealing as they can be operated also as an analog memory featuring a continuous range of metastable variable resistance states[20-23].

Metallic antiferromagnet CuMnAs is a material where several pioneering experiments connected with electrical switching of antiferromagnets were performed. In particular, at current densities

~ $10^6$ $Acm^{-2}$ a principally iso-thermal Néel vector reorientation by a field-like spin-orbit torque[5] was observed that can be detected by the anisotropic magnetoresistance (AMR), with relative resistance changes limited to tenths of a percent[6,21]. For higher current densities of ~ $10^7$ $Acm^{-2}$ switching of CuMnAs into highly resistive states occurs, with the relative resistivity change reaching 20% at room temperature and approaching 100% at low temperatures[23]. Unlike the spin-orbit torque reorientation of the Néel vector controlled by the angle or polarity of the switching current[6,7,21], this unipolar switching mechanism is independent of the writing current direction and can be achieved even for excitation by laser pulses[23]. Complementary imaging measurements, including X-ray magnetic linear dichroism photoemission electron microscopy (XMLD-PEEM) and magneto-Seebeck microscopy[23,24], have shown that the observed large switching signals were accompanied by a nanofragmentation of magnetic domains. This effect, occurring when the system is brought by the pulse to the vicinity of the Néel temperature, followed by a rapid quench into a metastable magnetically disordered state, is termed quench-switching[23]. Both these switching mechanisms were observed also in other antiferromagnetic metals – Néel vector reorientation in $Mn_2Au$ (Ref. [25,26]) and quench-switching in $Mn_2As$ (Ref. [27]), which shows their universal character. Consequently, these experiments were among the cornerstones establishing the antiferromagnetic spintronics[28-31].Despite this, the magnetic origin of these experiments was later questioned and attributed to thermal artifacts in patterned metal structures on substrates[32].

In this work, we experimentally separate, in a time domain with sub-ps resolution, resistance changes in CuMnAs devices due to a transient temperature increase and those due to the magnetic multi-level quench-switching effect. In particular, we show that the heat-related dynamics takes place on picosecond to hundreds of nanoseconds time scales, which is much faster than the quench-switched state relaxation dynamics that can be characterized by two main time components of ~ 10 ms and ~ 10 s in the Kohlrausch stretched exponential decay[23] (see Supplementary, Part C, for a detailed discussion about the possible origin of these two time components). Consequently, the heat-related dynamics can be used as a short-term memory where information about input stimuli, represented by femtosecond laser pulses in this work, is stored temporarily. When the switching threshold is reached, information is transferred to the device's variable multilevel resistance, which can serve as a long-term memory, where it can be retrieved at times up to $10^{15}$ longer than the input pulse duration. We also deduced the ~ 10 ps time scale as an upper estimate of time required for a conversion of the short-lived transient temperature increase to the long-lived metastable variable resistance states. Finally, we discuss the achieved results in a broader context of the research of compensated magnets towards analog logic-in-memory devices. In particular, we emphasize the unprecedented speed of the "leaky-sum" functionality in the studied CuMnAs-based memories and we also discuss their eventual applicability for a heat-based logic.

## RESULTS

### Samples and experimental setup

Epilayers of tetragonal CuMnAs with a thickness from 8 to 60 nm were grown by molecular beam epitaxy on a lattice matched GaP substrate[33]. Electron-beam lithography and wet chemical etching were used to pattern the devices[23]. (For more details on the sample preparation see Methods.) In the measurements, we employ a broad range of experimental techniques of ultrafast laser spectroscopy. In particular, to experimentally separate device resistance changes due to the temperature increase and quench-switching in CuMnAs (termed as "switching" for simplicity in the following), we use a stroboscopic optical-pump/electrical-probe method using a high-performance oscilloscope with a 6 GHz bandwidth and 150 fs laser pulses. To address dynamical processes in the studied devices in a very broad time ranges, from (sub)picoseconds to minutes, we combined two different varieties of one experimental method. As a first step, we performed a single-pulse experiment where the resistance change $\Delta R$ induced by a single femtosecond-laser pulse is detected by a high-performance oscilloscope, with an overall time-resolution of ~ 2 ns. As a second step, to study dynamics at (sub)picosecond to nanosecond timescales, we performed a double-pulse experiment. Here, the device is illuminated by a pair of laser pulses with an ultrashort mutual time delay and the time-resolution is given mainly by the time-width of the used femtosecond optical pulses. Moreover, we used all-optical pump-probe experiments to measure independently the electron-phonon relaxation time in CuMnAs and the heat dissipation from CuMnAs film to substrate, which strongly depends on the CuMnAs film thickness. (For more details on the experimental techniques see Methods and Supplemental Methods.)

Schematic of studied memory device is shown in Fig. 1a. After excitation by a single femtosecond laser pulse, the transient heating and rapid quenching of CuMnAs into nano-fragmented domain states[23,34] lead to a change of the device resistance. The dynamics of these two contributions can be separated by single-pulse and double-pulse experiments, which are shown schematically in Figs. 1b and 1d, respectively. The typical measured data in the single-pulse experiment are shown in Fig. 1c. In the inset of Fig. 1c we plot the resistance change $\Delta R$ detected over an extended time interval up to ~ 10 s after the impact of single femtosecond laser pulse using a non-stroboscopic method. It illustrates the two stretched-exponential relaxation components. The black rectangle highlights the temporal range dominated by the dynamics of the faster-relaxing component on which we focus in the systematic high-precision stroboscopic measurements. These are illustrated in the main panel of Fig. 1c for one laser fluence below the switching threshold and for two above-threshold fluences. The pulse-induced $\Delta R$ as a function of the time $t$ of the electrical probing are plotted relative to the device resistance at the end of the 10 ms measurement interval.

In Fig. 1e we plot the variable resistance detected at 0.1 ms after the incidence of a pair of laser pulses as a function of the time delay $\Delta t$ between the two successive pulses. We choose a laser fluence such that one pulse alone would not cause switching, while for a simultaneous incidence of the two pulses ($\Delta t = 0$) the combined fluence is safely above the switching threshold. In the following we analyze the observed phenomenology and the underlying physical mechanisms, starting from the single-pulse experiments.

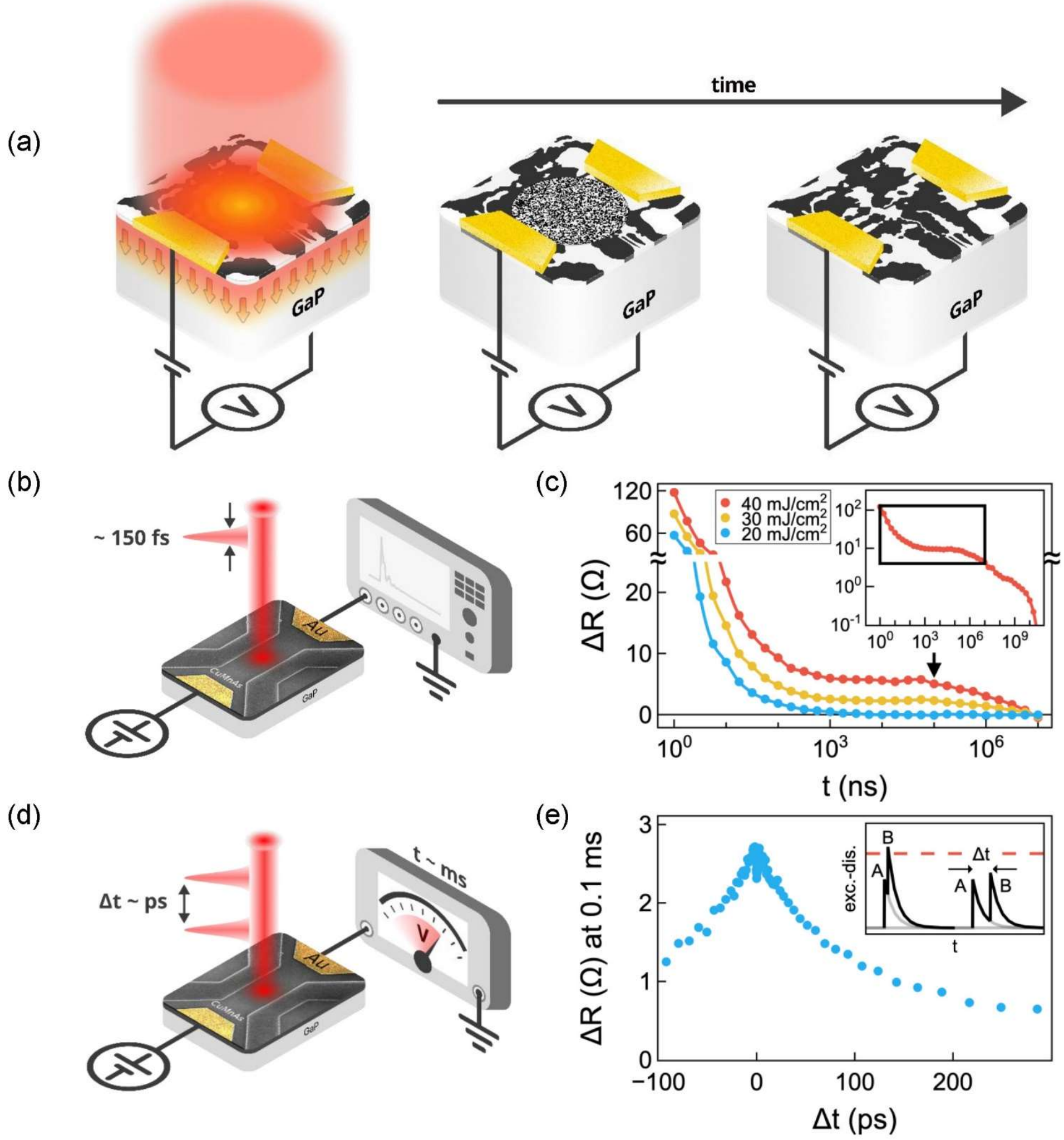

**Figure 1: Short-term and long-term memory in analog antiferromagnetic CuMnAs memory device. a,** Schematic of the memory device prepared from CuMnAs film with assumed fictional equilibrium magnetic domain structure (black and white features), which resembles the real domain structures in a film with a biaxial magnetic anisotropy[7,43]. Voltage is applied across two terminals (yellow patches). After excitation by a single femtosecond laser pulse (red column), heat dissipation from the CuMnAs layer to the GaP substrate (red arrows) plays the role of a leaky short-term memory. If the quench-switching threshold is reached[23], the device resistance increases due to the depicted assumed transient domain structure change (middle panel). At longer time scales, relaxation towards the equilibrium domain structure (panel on the right) plays the role of a leaky long-term memory. **b,** Schematic illustration of the setup for device optical excitation using a single femtosecond laser pulse, and for electrical readout of the resulting resistance change $\Delta R$ by a fast oscilloscope. **c,** Stroboscopic optical-pump/electrical-probe measurements for three fluences of the pump laser pulse in a device made from a 20-nm-thick CuMnAs epilayer. The pulse-induced $\Delta R$ as a function of the time $t$ of the electrical probing are plotted relative to the device resistance at the end of the 10 ms measurement interval. Inset: Non-stroboscopic measurement of $\Delta R$ relative to the equilibrium resistance, induced by a pump pulse with a fluence of 40 mJ/cm$^2$ over an extended time interval. The black rectangle highlights the temporal range dominated by the dynamics of the faster-relaxing component on which we focus in the main panel. **d,** Schematic of the experimental setup for device optical excitation by a pair of laser pulses with a mutual ultrashort delay time $\Delta t$. The resulting change in the device resistance is measured at macroscopic times $t$ after the incidence of the pair of laser pulses. **e,** Delay-time dependence of $\Delta R$ measured at $t = 0.1$ ms (depicted as an arrow in **c**) after the impact of the pair of laser pulses. Inset: Cartoon illustrating the excitation by two laser pulses ($A$ and $B$) followed by energy dissipation. In the double-pulse experiments, the fluence of an individual pulse is not sufficient to excite the CuMnAs antiferromagnet over the switching threshold (red dashed line). The pair of pulses causes switching and the resulting $\Delta R$ depends on the delay time between the pulses.

**Single-pulse experiments**

Data shown in Fig. 2 illustrate our decomposition of the measured time-dependence of $\Delta R$ into a heating effect (independent of the switching), and the switching effect. Fig. 2a shows the as-measured fluence-dependence of $\Delta R$ at three distinct probing times $t$ after the incidence of a single laser pulse. The linear dependence below the switching-threshold fluence corresponds to the heating contribution. The solid lines in Fig. 2a then extrapolate the linear heating contribution also to fluences above the switching threshold. In Fig. 2b we plot the fluence-dependence of $\Delta R$ after subtracting the linear heating contribution. Data for $t$ = 1 μs and 1 ms are multiplied by the indicated factors to highlight the universal switching-threshold characteristics. The continuous increase of the device resistivity with the pulse fluence is connected with an increased amount of the CuMnAs material that was switched to the higher resistive state (see Fig. S2a). The line in Fig. 2b is a fit based on a model that considers the experimental Gaussian spatial profile of the laser beam, and describes the threshold fluence $F_{TH}$ and the scaling of the switching contribution to $\Delta R$ with the volume of the antiferromagnet excited above the switching threshold. (For more details see Supplemental Note S1.) Because of the Gaussian profile of the laser beam, we do not systematically explore in our optical-pulsing experiments the high-fluence regime corresponding to saturated $\Delta R$. In this regime, the excitation at the center of the laser beam exceeds the threshold for structurally damaging the antiferromagnetic crystal.

In Figs. 2c and 2d we plot, respectively, the $t$-dependence of the heating and switching contributions to $\Delta R$. They highlight that the switching contribution to $\Delta R$ can be reliably extracted using our electrical probing method starting from ~ 10 ns after the laser pulse. In this time range, the heat-related signals decay with time constants $\tau_2$ ~ 10 ns and $\tau_3$ ~ 100 ns, which we attribute to a heat transfer from the GaP substrate to the sample holder and environment. Importantly, the switching-related signals do not decay up to ~ 100 μs, when the fastest component in the Kohlrausch stretched-exponential model starts to be apparent (see Supplemental Note S2).

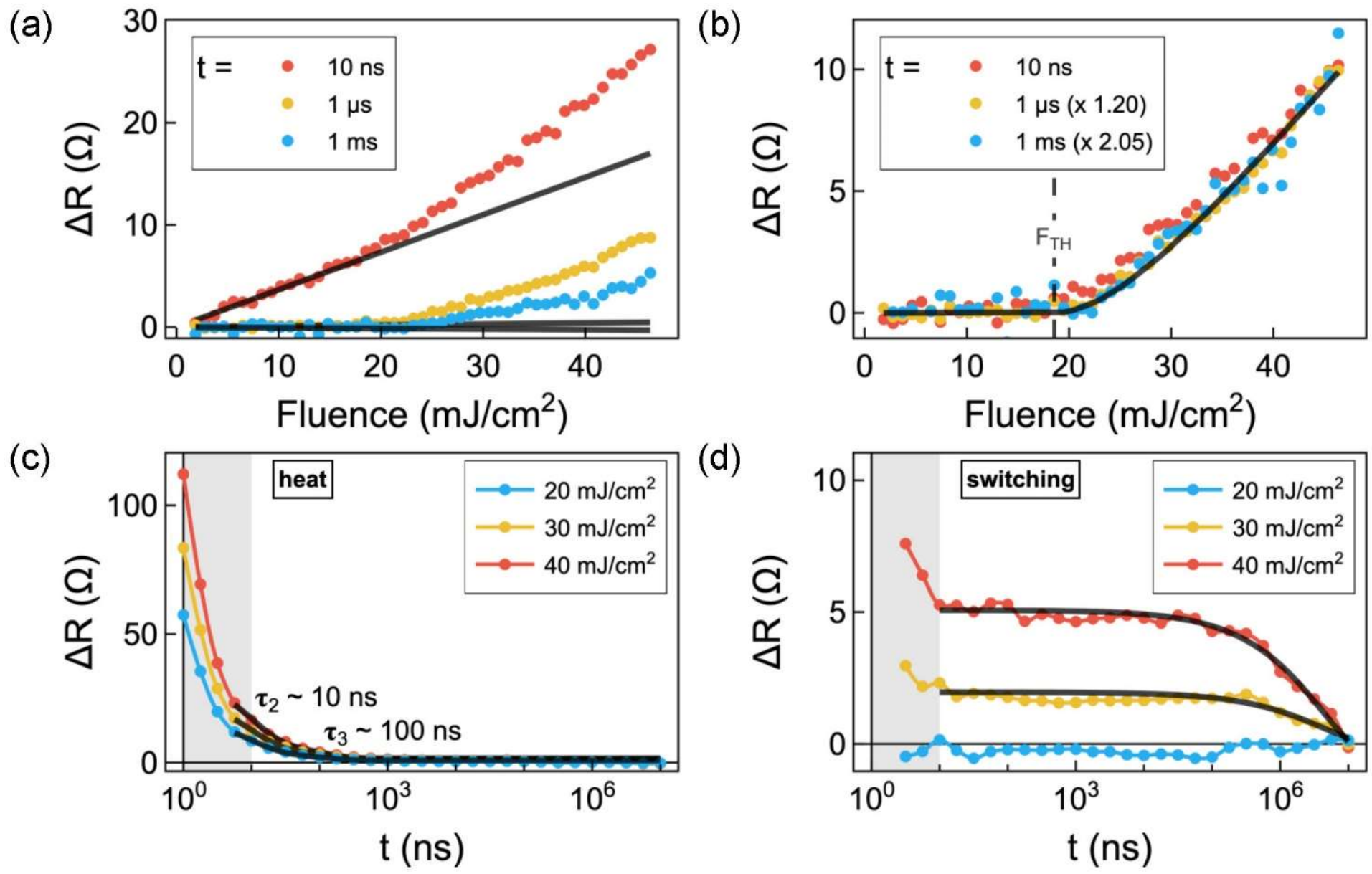

**Figure 2: Decomposition of measured device resistance change $\Delta R$ into a heating contribution and a switching contribution in a single-pulse experiment.**

**a,** As-measured fluence dependence of $\Delta R$ at three distinct times $t$ after the incidence of a single laser pulse (dots) in a device made from a 20-nm-thick CuMnAs epilayer. The heating contributions to the signals are depicted by solid lines. **b,** Same as **a** with the heating contributions subtracted; the obtained values (dots) for $t$ = 1 μs and 1 ms are multiplied by the indicated factors to show the common fluence dependence. The line is a fit using a theoretical threshold model (see main text and Supplemental Note S1) that defines the value of the fluence threshold $F_{TH}$. **c, d,** Deduced time evolution of heating and switching contributions to $\Delta R$, respectively. Gray areas denote the time window where the applied decomposition method becomes unreliable (see Supplemental Notes S2 and S3 for a detailed discussion). The indicated heat relaxation time constants $\tau_2$ and $\tau_3$ were obtained by a double exponential fit (black solid lines) of experimental data in **c**. The relaxation of the switching signal (see Supplemental Note S2) in **d** is characterized by the Kohlrausch stretched-exponential model (lines).

## Double-pulse experiments

We now proceed with the analysis of the phenomenology and underlying physics of the double-pulse experiments, which are aiming to reveal the heat-related dynamics at (sub)picosecond timescales. Here, the device is illuminated every 10 ms by a pair of laser pulses with an ultrashort mutual time delay $\Delta t$ (see Fig. S5) and the resulting device resistance change is measured at macroscopic times $t$, typically at $t \approx 0.1$ ms, after the incidence of the double-pulse as a function of the delay time $\Delta t$ between the two pulses. In Figs. 3a and 3b we compare measurements of $\Delta R$ in devices fabricated from a 20 nm and 50 nm thick CuMnAs films. Initially, for $\Delta t$ on a ~ ps scale, $\Delta R$ increases with increasing $\Delta t$ (see the insets). This is then followed by decreasing $\Delta R$ on $\Delta t$-scales of ~ 100 ps in the 20 nm film and ~ 1000 ps in the 50 nm film.

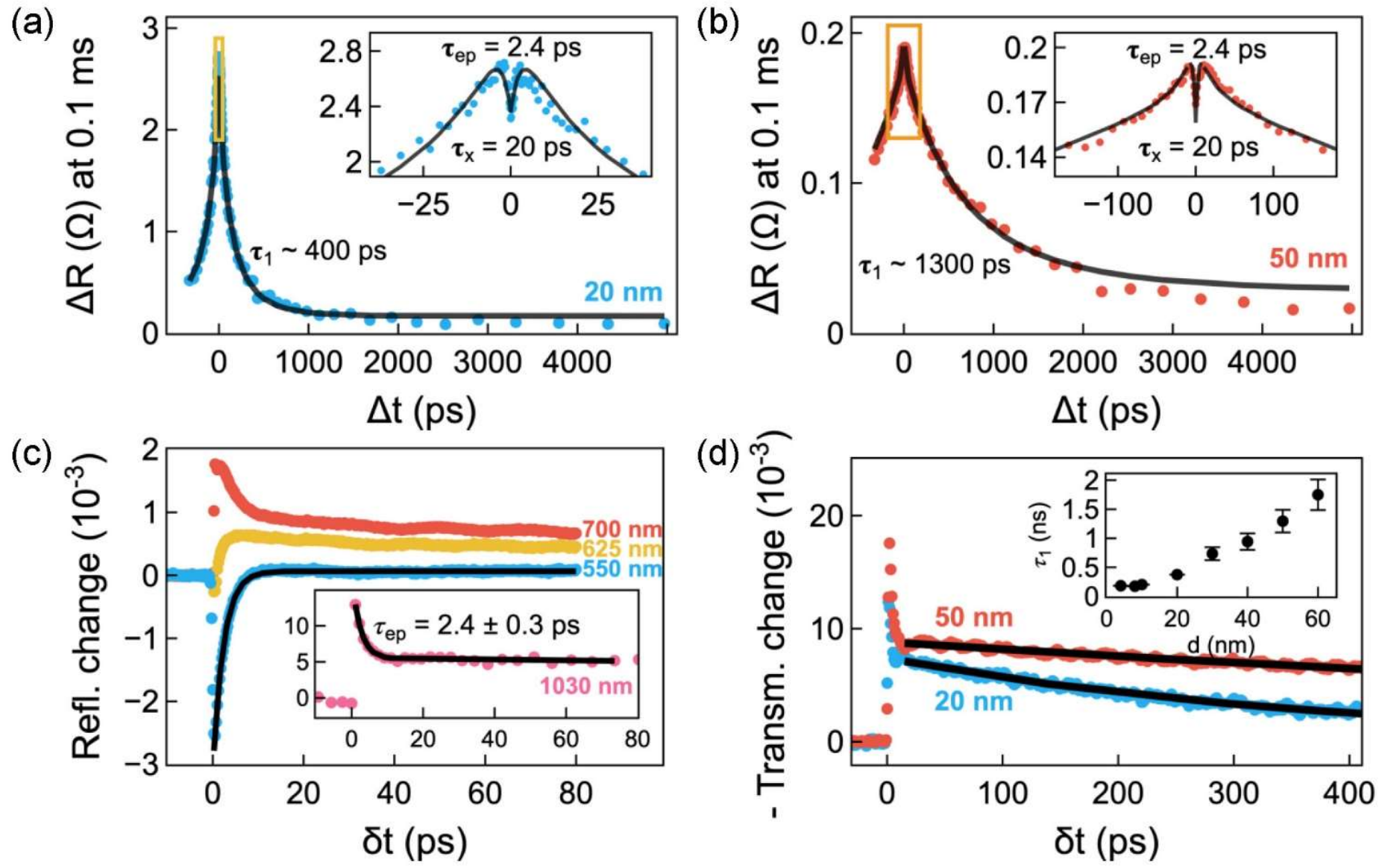

**Figure 3: Delay-time dependence of $\Delta R$ in a double-pulse experiment. a,** Delay-time $\Delta t$ dependence of $\Delta R$ measured 0.1 ms after the incidence of the pair of laser pulses, which is indicated by a black vertical arrow in Fig. 1c, in a device made from a 20-nm-thick CuMnAs epilayer (dots). The solid line is a model described in the main text and Supplemental Note S4 with electron-phonon relaxation time $\tau_{ep}$, time constant $\tau_x$ and thermal-relaxation time $\tau_l$ as fitting parameters. Inset: Detailed view of the data region near the zero-time overlap of the pulse pair, illustrated as a rectangle in the main panel. **b,** Same as **a** for a device made from a 50-nm-thick CuMnAs epilayer. **c,** Dynamics of differential reflectivity $d\mathscr{R}/\mathscr{R}$ in 50-nm-thick epilayer at 15 K for three selected wavelengths of probe pulses after excitation by 820 nm pump pulse with a fluence of 3 mJ.cm$^{-2}$ (dots). Inset: Dynamics of $d\mathscr{R}/\mathscr{R}$ measured at 300 K using pump and probe pulses with a wavelength of 1030 nm and pump fluence of 20 mJ.cm$^{-2}$. Lines are fits yielding the electron-phonon relaxation time $\tau_{ep}$ (see main text and Supplemental Note S4 for detailed discussion). **d,** Laser-pulse-induced dynamics of transient decrease of differential transmission $d\mathscr{T}/\mathscr{T}$ measured by optical-pump/optical-probe experiment in 20-nm and 50-nm-thick CuMnAs epilayers at 15 K using pump and probe pulses with a wavelength of 820 nm (dots) and pump fluence of 3 mJ.cm$^{-2}$. Lines are fits by a mono-exponential decay function with a characteristic time constant $\tau_l$ corresponding to the heat dissipation from the CuMnAs epilayer to the GaP substrate. Inset: Dependence of $\tau_l$ on the epilayer thickness; the error bars depict the experimental uncertainty of $\tau_l$ determination from the pump-probe experiment.

A physical interpretation of the initial increase of $\Delta R$ is suggested by ultra-fast transient reflectivity measurements of CuMnAs, shown in Fig. 3c, obtained using a stroboscopic optical-pump/optical-probe method. In the inset of Fig. 3c, we show the optical-pump/optical-probe measurement using the same wavelength of pump and probe pulses as that in the double-pulse experiments in Figs. 3a and 3b, and using a sub-threshold pump fluence. We observe that the pump pulse causes an abrupt increase of the reflectivity that decays with a characteristic time constant $\tau_{ep} = 2.4 \pm 0.3$ ps. Based on the non-degenerate pump-probe measurements over a range of probe wavelengths shown in the main panel of Fig. 3c, we interpret[35-37] this time constant as the electron-phonon relaxation time in CuMnAs[38]. We, therefore, attribute the initial increase of $\Delta R$ with increasing $\Delta t$ on the ~ ps scale to the abrupt increase of the CuMnAs reflectivity caused by the first pulse, which is influencing the switching efficiency of the second pulse.

In Fig. 3d we show transient transmission measurements using the optical-pump/optical-probe method over a larger range of pump-probe delays reaching several hundred ps. The relaxation of the transmission on these longer time scales reflects the heat dissipation from the CuMnAs film[38] and the inferred thermal-relaxation time $\tau_1$ increases with increasing film thickness, as highlighted in the inset of Fig. 3d. Consistently, the decay of $\Delta R$ with $\Delta t$ is considerably slower in the 50 nm film than in the 20 nm film. Qualitatively, we can therefore attribute the ~ 100 − 1000 ps scale relaxation of $\Delta R$ in Figs. 3a and 3b to the dissipation of the excitation energy delivered by the first pulse until the arrival of the second pulse. Interestingly, in all-optical pump-probe experiments only time constants $\tau_{ep}$ and $\tau_1$ are apparent (see inset of Fig. 3c). In contrast, an additional time constant $\tau_x$ ~ 20 ps is present in the measured decrease of $\Delta R$ with $\Delta t$, as shown in insets of Figs. 3a and 3b. This time constant seems to be independent on the CuMnAs film thickness but its physical origin is not clear at this moment (see Supplemental Notes S3 and S4 for a detailed analysis and discussion of double-pulse experiments).

**Evaluation of characteristic times from single and double-pulse experiments**

To quantitatively infer the thermal-relaxation time $\tau_1$ from the double-pulse experiments, we introduce a concept of effective switching fluence $F_{eff}$. This quantity represents the effective energy delivered by the pulse-pair to the CuMnAs epilayer by taking into account the partial dissipation of heat from the first pulse before the arrival of the second pulse. Assuming a monoexponential heat dissipation (Fig. 3d), $F_{eff}(\Delta t) = F_1 \exp(-\Delta t/\tau_1)+F_2$, where $F_{1(2)}$ are the incident fluences of the first (second) pulse (see Supplemental Note S4 for a detailed discussion). From the experimental point-of-view, this quantity represents the fluence of a single laser pulse that would generate the same $\Delta R$ signal as a laser pulse-pair with a given time delay $\Delta t$. To convert the measured dependence $\Delta R(\Delta t)$ to $F_{eff}(\Delta t)$ one can use the experimentally measured intensity dependence $\Delta R(F)$ in the single-pulse experiment, depicted in Fig. 2b (see Supplemental Note S4.1 and Fig. S7, for details). The result for the 20 nm CuMnAs film is shown in Fig. 4a with the inferred value $\tau_1$ ~ 400 ps fully consistent with the thermal-relaxation time obtained from the transient transmission measurement shown in Fig. 3d.

Due to the capability of our simple theoretical model to consistently describe all our experimental data, we can also use it to gain deeper insight into the time scales involved in the CuMnAs switching to high resistivity states[23]. From an expression, $F \exp(-t_{\mathrm{above}}/\tau_1) = F_{\mathrm{TH}}$, and the experimental values of $\tau_1$ and the switching-threshold fluence $F_{\mathrm{TH}}$ (see Fig. 2b), we can estimate the time $t_{\mathrm{above}}$ the antiferromagnet remains excited above the switching threshold after a pulse of fluence $F$ at a spatial position corresponding to the center of the Gaussian laser beam. As shown by points in Fig. 4b, we can experimentally resolve switching signals for which $t_{\mathrm{above}}$ approaches the ~ 10 ps scale. This provides an upper estimate of the time required for converting the short-lived transient temperature increase to long-lived metastable variable resistance states. This timescale is crucial for in-memory logic operations based on the transfer from short-term to long-term memory.

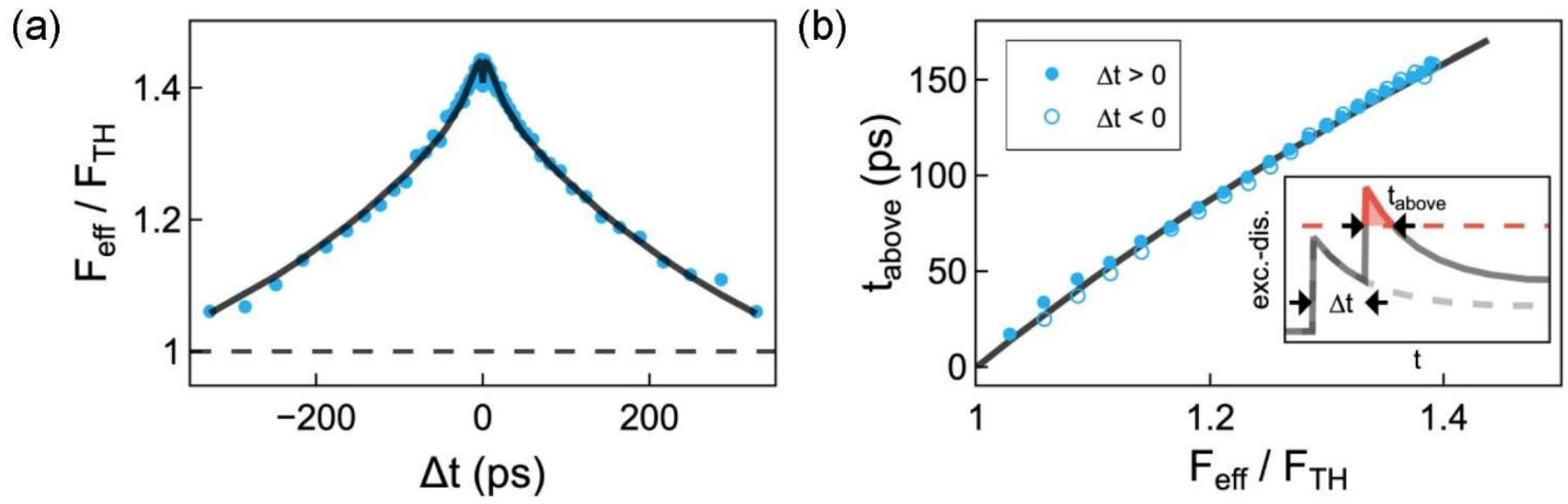

**Figure 4: Evaluation of characteristic times from double-pulse experimental data. a**, Delay-time dependence of $\Delta R$ measured 0.1 ms after the incidence of the pair of laser pulses in a device made from a 20-nm-thick CuMnAs epilayer (depicted in Fig. 3a) expressed as an effective switching fluence $F_{eff}$, which represents the fluence required for a single pulse to induce the same $\Delta R$ as the pair of pulses; the obtained fluences are normalized by the switching-threshold fluence $F_{TH}$ (dots), line is a fit with $\tau_1 \approx 400$ ps (see main text and Supplemental Note S4 for details). **b,** Time $t_{\mathrm{above}} = \tau_1 \ln(F/F_{TH})$, describing how long the antiferromagnet remains excited above the switching threshold after the laser pulse impact, as a function of the pulse fluence inferred from data shown in part **a**. Inset: Illustration of $t_{\mathrm{above}}$ in the excitation-dissipation cartoon with the threshold depicted by a red dashed horizontal line.

## DISCUSSION

We discuss our experimental results within the broader context of the research of compensated magnets towards ultra-scalable digital memory devices and analog logic-in-memory devices. Components for artificial spiking neural networks are among the prominent applied-oriented research areas based on unconventional functionalities in antiferromagnets[20-23,39-41]. Earlier experimental works have focused on spiking neuromorphic functionalities in antiferromagnetic devices operating down to ~ 100 ns time scales[39]. Our results open new possibilities for implementing complex time-dependent logic-in-memory functionalities.

In our devices, the short-term memory functionality is based on ultrafast heat dynamics in nanometer-thin CuMnAs films, which is generic to many materials. However, the favorable feature of CuMnAs is its ability to be switched by accumulated heat to a continuous range of metastable variable resistance states, with relative resistance-changes reaching ~ 20% at room temperature and ~ 100% at low temperatures[23,42]. Previous studies ascribed these higher-resistance states to a formation of unconventional nano-scale magnetic textures in the compensated antiferromagnet[23]. Magnetic textures in CuMnAs and their modification by applied current pulses have been studied by microscopy methods based on X-ray magnetic linear dichroism[6,7,43-45] and thermoelectric effects[24]. The isothermal spin-orbit torque switching by means of 90° domain-wall motion was directly visualized by XMLD-PEEM[7], complementing the electrical AMR detection. Moreover, measurements in the spin-flop (reorientation) fields of ~ 2 T confirmed that AMR associated with the reorientation of the Néel vector is on the ~0.1% scale in CuMnAs (Ref. 46). In contrast, the resistance changes induced by the quench-switching are not affected significantly by magnetic field (Ref. 42) and the enhanced resistivity is attributed to scattering off the strong (metastable) magnetic disorder. Unlike the reorientation switching which is non-volatile, the thermally induced quench switching signal relaxes towards the low-resistivity equilibrium magnetically uniform state. A possible form of the strong magnetic disorder are atomically sharp 180° domain walls, which were directly observed in CuMnAs by scanning transmission electron microscopy with spin and atomic spatial resolution[34]. The magnetic origin of the quench-switching signal was further confirmed in recent experiments in isostructural antiferromagnet $Mn_2As$, where the Néel temperature is 100 degrees higher than in CuMnAs (Ref. 27). Importantly, the threshold switching current and the relaxation time scale with the Néel temperature in these two antiferromagnets[27].

The working principle of our CuMnAs-based memory devices resembles phase-change memories (PCMs), where switching between amorphous (high resistivity) and polycrystalline (low resistivity) states is also controlled by delivered heat[11,47]. PCMs are a rather mature material system where many different optical-based in-memory computing functionalities[12,13,15] were already demonstrated, including their integration to photonic memory devices and integrated circuits[12,48]. The speed limit of PCMs is set by the crystallization time, which has been reduced from milliseconds[49] to sub-nanoseconds[50] through decades of effort, including preparation of nanometer-sized devices. At present, the intrinsic time scale characterizing the formation of higher-resistance states in CuMnAs is not known. Our experimental results, without any device and/or material optimization, show that the characteristic time is shorter than ~ 10 ns, which is the time resolution of our single-pulse high-performance oscilloscope-based experiment (see Fig. 2d). Moreover, the performed analysis reveals an *upper estimate* of ~ 10 ps for the time required for a conversion of the short-lived transient temperature increase to the long-lived metastable states in CuMnAs. Overall, this points that antiferromagnetic CuMnAs-based memories have a perspective to become much faster than PCMs, which is in accord with much faster magnetization dynamics in compensated magnets compared to the recrystallization dynamics utilized in PCMs. We expect that the concept of heat-based short-term memory combined with magnetic-ordering-based long-term memory should be applicable in many other magnetic materials where the corresponding ordering temperature would play the role of the characteristic threshold temperature. For example, heat-based switching was reported in GdFeCo (Ref. [51]), FeRh (Ref. [52]), $Mn_3Sn$ (Ref. [53]) and $Mn_5Si_3$ (Ref. [54]).

CuMnAs-based memories exhibit the ”leaky-sum” functionality, resembling the neuron-like non-linear response. In previous studies of CuMnAs devices, the signal summation was performed at time scales ranging from milliseconds to minutes[22,23], corresponding to the characteristic time components in the relaxation dynamics of higher-resistance states, which serve as a long-term memory. In contrast, in this paper we concentrate on the ”leaky-sum” functionality on picosecond to hundred nanosecond time scales, utilizing the heat-related dynamics as a short-term memory (see also Supplemental Figs. S5a and S5b). These ultrashort time scales are unprecedented also in the context of neuromorphic artificial synaptic devices, where the short-term to long-term memory transition has been extensively studied[55-61], but only at millisecond and longer time scales. This combination of heat-based short term memory with quench-switching-based long term memory enabled us to demonstrate experimentally[62] the ultrafast version of several operations commonly used in neural networks, including temporal latency coding of a grayscale image into sub-nanosecond pulse delays and pulse-pattern recognition.

In this study, we use light pulses as ultrashort input stimuli for our devices. The main reason is that the generation of femtosecond laser pulses is a mature technology readily available in commercial table-top laser systems and it is the fastest stimulus experimentally available. However, as metals are absorbing in a broad frequency range, THz pulses[22], electrical pulses[6,21,23], and/or their combinations with optical pulses[23] can be also used for switching the CuMnAs-based devices. The required energy-per-pulse for achieving the switching is primarily determined by the device size, i.e., by the switched volume of the antiferromagnet (see Supplementary Fig. S2). In this paper, we used microbar devices with dimensions $10 \times 20$ μm$^2$ illuminated by Gaussian spatial-profile laser pulses with a diameter of 20 μm, where the switching threshold fluence corresponds to 36 nJ of delivered energy per pulse (see Supplemental Note S1). For smaller devices (illuminated by more tightly focused laser beams) the required energy is considerably smaller. We studied devices with sizes down to ~ 1 $\mu m^2$, where the switching threshold energy reaches ~ 100 pJ (see inset in Fig. 4a in Ref. 62 and the discussion therein). Considering the nanometer-sized domains[43] and the recent discovery of atomically-sharp domain walls[34] in CuMnAs antiferromagnet, it is reasonable to expect that the device size could be potentially scaled down to ~ nm dimensions. At this limit, assuming a CuMnAs device with a volume of 10 $nm^3$, the switching energy would be ~ 3.5 fJ, which is comparable to the ~ 1-100 fJ per synaptic event in biological systems[63].

The measured heat-dissipation dynamics in our devices comprise time constants ranging from ~ 20 ps to 100 ns. Consequently, this heat-based short term-memory can be exploited for any applicable excitation stimuli (i.e., light pulses[23], THz pulses[22] or electrical pulses[6,21,23]) with a duration shorter than ~ 100 ns. Moreover, for excitation of CuMnAs by optical pulses the available playground is considerably richer due to a presence of a transient change of the sample reflectivity, which determines what fraction of incident laser pulse energy is transferred to the heat pulse that is inducing the actual CuMnAs switching. If the wavelength of laser pulses is 1 030 nm, as in the case of the double-pulse experiment depicted in Figs. 3a and 3b, the abrupt increase of the CuMnAs reflectivity caused by the first pulse (see inset in Fig. 3c) is *reducing* the switching efficiency of the second pulse. This is analogous to the pair-pulse depression (i.e., a transient synaptic strength decrease due to the neurotransmitter depletion) in a biological synapse[64-67]. Instead, if the second laser pulse wavelength were 550 nm, for example, the transient reduction of the CuMnAs reflectivity (Fig. 3c) would lead to an *enhanced* sensitivity to the second pulse, which resembles the pair-pulse facilitation (i.e., a transient synaptic strength increase) instead. Moreover, as this effect is governed by the electron-phonon relaxation, with a characteristic time constant of 2.4 ± 0.3 ps in CuMnAs, it enables to change the short-term memory properties, resembling the short-term synaptic plasticity, on a picosecond time scale, which is quite unique – see Fig. 3 in Ref. 65 where the current status in the field of short-term artificial synaptic devices is reviewed.

Finally, we address the achieved results from the view-point of heat-based logic[68-70]. Heat and its dissipation are generally regarded as nuisance in electronics where they are, for example, limiting the processor frequency. Nevertheless, heat can be also used as an input in thermal computing systems, which are based on a dynamical control of heat flows. The corresponding thermal circuits are envisioned to be powered by excess heat in integrated circuits, making it a rather promising platform thanks to its low energy consumption. At present, there are several proposed theoretical concepts of heat-based computation[68,71,72] and also some experimental demonstrations of thermal transistors[73-75]. To be compatible with a mainstream electronics, the thermal logical values have to be converted to electrical ones at some point. For this purpose, temperature-biased normal-metal/ferromagnetic-insulator/superconductor tunnel junctions[71] can be used, for example. Our experiment show that this thermal-to-electrical signal conversion can be performed also in the studied CuMnAs-based memory devices. In particular, these devices can sum the incident short thermal pulses and change their long-term electrical resistance when temperature reaches the quench-switching threshold, which makes these devices potentially interesting also for this research direction.

## METHODS

### Samples

Tetragonal CuMnAs epilayers were grown using molecular beam epitaxy on lattice-matched GaP substrate at temperatures around 200 °C. The films were capped with a 3 nm Al layer, which quickly oxidized after removal from the vacuum, preventing oxidation of the CuMnAs. Microbar devices (Fig. 1b) with typical dimensions of $10 \times 20\ \mu m^2$ were lithographically patterned from 20- or 50-nm-thick epilayers (with a sheet resistance of 50 and 20 Ω, respectively). Electron-beam lithography and wet chemical etching were used to pattern the devices. The aluminum cap was removed in 2.7% $C_4H_{13}NO$ (for 3 s) and a mixture of $H_3PO_4$:$H_2O_2$:$H_2O$ in ratio 1:10:400 was used as the CuMnAs etchant, providing an etching rate of 50 nm per 20 s. A custom-designed, two-terminal high-frequency printed circuit board sample holder was employed for bonding gold contacts sputtered onto the opposing ends of the microbars.

### Experiment combining optical writing and electrical readout

An Yb-based femtosecond laser system (Pharos, Light Conversion) was used as the light source for optical writing. Laser pulses, with a 150-fs pulse duration, a central wavelength of 1030 nm, and a user-selectable repetition rate, were divided into two branches with a precise control of their mutual time delay (see Supplementary Fig. S1a). The laser pulses were focused onto a single spot on the sample, exhibiting a Gaussian intensity profile with a diameter of 20 μm (see Supplemental methods). The fluence of laser pulses in each branch was independently controlled by a computer, enabling fluences up to 50 mJ/cm².

Time-resolved electrical readout was performed using a Rohde & Schwarz RTP064 high-performance oscilloscope, with a 6 GHz bandwidth, while a Rigol DG1000Z served as the voltage source. The electrical setup schematic is illustrated in Supplementary Fig. S1b. Voltage measurements were taken across the input resistance of the oscilloscope, which was connected in series with the sample. The sample's resistance was deduced from these voltage readings (see Supplemental Methods), and the resistance increase, relative to the device resistance at the end of the measurement time window, was computed. Typically, 2.5 million data points were collected during a single acquisition of the selected measurement range. To reduce a noise at high frequencies without compromising the time resolution, the majority of experiments were performed with a laser working at a 100 Hz repetition rate, corresponding to 10 ms time-spacing between adjacent laser pulses, and ~ 100 acquisitions were averaged for each measurement. Consequently, only time transients faster than 10 ms are apparent in the measured data [see Fig. 1(c)]. All measurements were conducted at room temperature.

Using this experimental setup, two types of experiments can be performed, which are shown schematically in Figs. 1b and 1d, respectively. In the first type, the sample is illuminated by a single laser pulse every 10 ms, and the oscilloscope records the corresponding resistance changes within this time window. Here, the time variable $t$ represents the oscilloscope measurement time, with $t = 0$ ns defined as the time when the maximum value of the measured waveform is obtained (see Fig. S5a). The overall time-resolution of this oscilloscope measurement is visible in the ~ 2 ns rising time of the electrical signal after the femtosecond laser pulse impact, which is depicted as a gray area in Fig. S5a. In the second type, the sample is illuminated every 10 ms by a pair of laser pulses with a mutual ultrashort delay time $\Delta t$ (see Fig. S5) and the resulting change in the device resistance is measured at macroscopic times $t$, typically at $t \sim 0.1$ ms, after the incidence of the double-pulse. The time-zero delay, $\Delta t = 0$ ps, is defined by the simultaneous incidence of both laser pulses on the device. Importantly, this second experimental scheme provides a much better time-resolution, which is given mainly by the time-width of the used optical pulses of ~ 150 fs.

**Pump-probe experiment**

We performed both a degenerate pump-probe experiment, where the same wavelength of pump and probe pulses is used, and also its non-degenerate variety, with a non-equal wavelength of pump and probe pulses. For the degenerate pump-probe experiment at 1030 nm, we used the Yb-based femtosecond laser system and experimental parameters described above. In addition, a degenerate pump-probe experiment at 820 nm was performed using a femtosecond Ti:Sapphire oscillator (Mai Tai, Spectra Physics) generating ≈ 150 fs laser pulses at a repetition rate of 80 MHz – see Fig. 1a in Ref. 76 for the corresponding experimental setup.

For the non-degenerate pump-probe experiment, we used the fundamental output from the Ti:Sapphire oscillator at 820 nm as a pump beam and the output from an optical parametric oscillator (Inspire, Spectra Physics), tunable from 550 nm to 700 nm, as a probe beam. The pulses were focused onto the same spot (with a Gaussian diameter of ≈ 35 µm) on the sample at nearly normal incidence (see Supplemental Methods). The laser fluence of the pump pulses was approximately 3 $mJ/cm^2$ (i.e., well below the CuMnAs switching threshold), while the probe pulses were at least 50 times weaker. The experiment was performed at temperature of 15 K. Here, we simultaneously measure the pump-induced transient changes of CuMnAs transmission (the differential transmission $d\mathscr{T}/\mathscr{T} = (\mathscr{T}_E - \mathscr{T})/\mathscr{T}$, where $\mathscr{T}_E$ and $\mathscr{T}$ are transmissions with and without the pump pulse, respectively) and reflectivity (differential reflectivity $d\mathscr{R}/\mathscr{R} = (\mathscr{R}_E - \mathscr{R})/\mathscr{R}$, where $\mathscr{R}_E$ and $\mathscr{R}$ are reflectivities with and without the pump pulse, respectively). For more details, see Ref. 38.

Further experimental details can be found in Supplemental methods.

## RESOURCE AVAILABILITY

### Lead contact

Requests for further information and resources should be directed to and will be fulfilled by the lead contact, Petr Němec (petr.nemec@matfyz.cuni.cz).

### Materials availability

This study did not generate any reagents.

**Data and code availability**

- Data reported in this paper has been deposited at Zenodo at [10.5281/zenodo.14856616] and are publicly available as of the date of publication.
- This paper does not report original code.
- Any additional information required to reanalyze the data reported in this paper is available from the lead contact upon request.

**ACKNOWLEDGEMENTS**

This work was supported by TERAFIT project No. CZ.02.01.01/00/22_008/0004594 funded by Ministry of Education Youth and Sports of the Czech Republic (MEYS CR), programme Johannes Amos Comenius (OP JAK), call Excellent Research. The authors acknowledge funding by the Czech Science Foundation (grant no. 21–28876J), by ERC Advanced Grant no. 101095925, by the Grant Agency of the Charles University (grants no. 166123 and SVV–2024–260720), by CzechNanoLab Research Infrastructure supported by MEYS CR (LM2023051), and by MEYS CR project LNSM-LNSpin.

**AUTHOR CONTRIBUTIONS**

P.N., K.O., M.S., J.Z., L.N., and T.J. planned the experiments, V.N. and F.K. prepared the samples, M.S., J.Z. and F.T. prepared the software for the data acquisition, M.S., J.Z., A.F., and P.K. performed the experiments, T.J., M.S. and P.N. wrote the manuscript with contributions from all authors.

**DECLARATION OF INTERESTS**

The authors declare no competing interests.

**SUPPLEMENTAL INFORMATION**

Supplemental materials consist of Supplemental methods (containing Figure S1), Note S1. Switching threshold model and effect of local resistivity change on device resistance (containing Figure S2), Note S2. Single-pulse method: Separation of switching and heat contributions from measured resistance change (containing Figures S3 and S4), Note S3. Double-pulse method: Heat accumulation and separation of signals for two mutually-time-delayed pulses (containing Figures

S5 and S6), Note S4. Heat dissipation dynamics (containing Figures S7, S8 and S9, Table S1), and Supplemental references.

# Picosecond transfer from short-term to long-term memory in analog antiferromagnetic memory device: Supplementary information

Miloslav Surýnek[1,4], Jan Zubáč[2,1,4], Kamil Olejník[2], Andrej Farkaš[2,1], Filip Křížek[2], Lukáš Nádvorník[1], Peter Kubaščík[1], František Trojánek[1], Richard P. Campion[3], Vít Novák[2], Tomáš Jungwirth[2,3], and Petr Němec[1,5*]

[1]Faculty of Mathematics and Physics, Charles University, Ke Karlovu 3, 121 16 Prague 2, Czech Republic
[2]Institute of Physics ASCR, v.v.i., Cukrovarnická 10, 162 53 Prague 6, Czech Republic
[3]School of Physics and Astronomy, University of Nottingham, Nottingham NG7 2RD, United Kingdom
[4]These authors contributed equally

[5*]Lead and corresponding author: petr.nemec@matfyz.cuni.cz

CONTENTS

**Supplemental methods**

The stroboscopic optical writing experiments were performed using an Yb-based femtosecond laser system (Pharos, Light Conversion). As shown in Fig. S1a, laser pulses, with a pulse duration of 150 fs, a central wavelength of 1030 nm, and a user-tunable repetition rate, were split into two optical paths via a polarizing beam splitter (PBS), where their intensity was independently regulated by a pair of computer-controlled half-waveplates (λ/2). The laser pulses were subsequently focused into a single spot on the sample, producing a Gaussian intensity profile

$$I(r) = I_0 \cdot e^{-\frac{r^2}{w^2}} \quad \text{(SM.1)}$$

where $r$ is the radial distance from the center of the Gaussian profile. The experimentally achieved beam size, characterized by a full-width $2w \approx 20$ µm, was determined using the knife-edge method.

The corresponding time-resolved electrical measurements were carried out using a Rohde & Schwarz RTP064 high-performance oscilloscope with a bandwidth of 6 GHz for readout and a Rigol DG1000Z as a voltage source ($U_{\text{source}}$ = 150 mV, $R_{\text{source}}$ = 50 Ω). The schematic of the electrical setup is presented in Fig. S1b. We recorded the voltage $U$ across the input resistance $R_{\text{scope}}$ = 50 Ω of the oscilloscope, connected in series with the sample. The resistance of the sample was then determined by

$$R = \frac{U_{\text{source}} * R_{\text{scope}}}{U} - R_{\text{source}} - R_{\text{scope}} \quad \text{(SM.2)}$$

Subsequently, we corrected $R$ by applying the oscilloscope's attenuation factor (in our case *atten* = 1 dB):

$$R_{\text{sample}} = R/\sqrt{10^{atten[dB]/10}} \quad \text{(SM.3)}$$

Typically, we collected 2.5 M data points during a single acquisition over a 10 µs or 10 ms measurement range, corresponding to the laser's duty cycle. To minimize noise at high frequencies without compromising time resolution, we generally averaged curves from approximately 100 acquisitions.

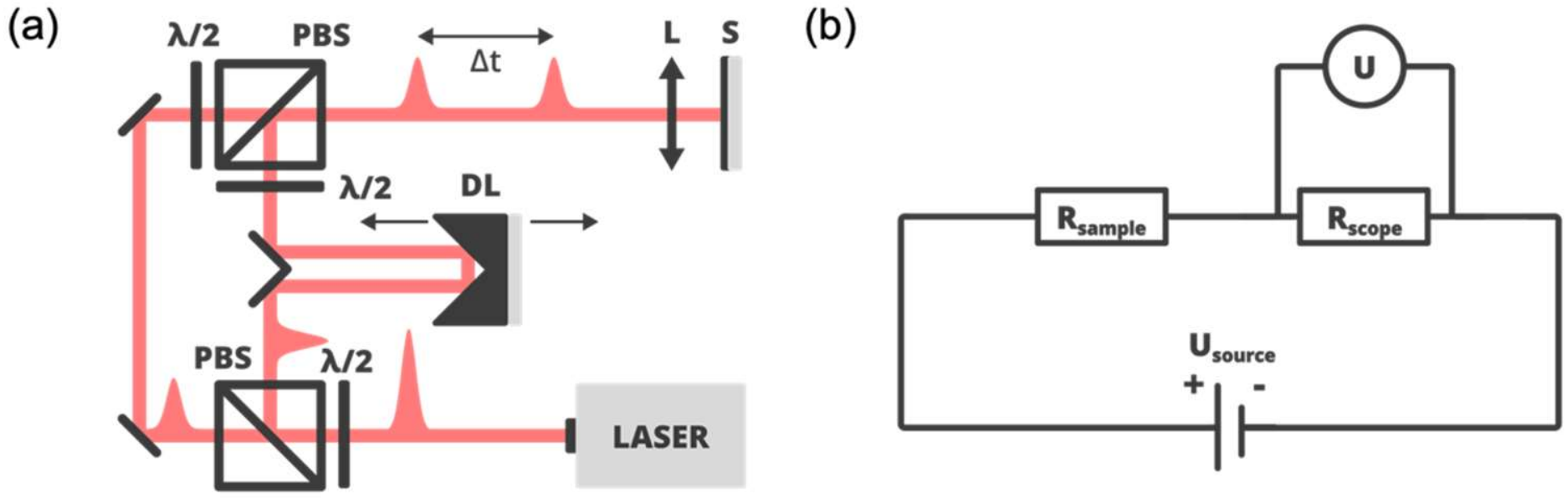

**Fig. S1 | Schematic illustration of the experimental setup for optical switching measurements with electrical readout. a,** Optical setup for stroboscopic experiments: A beam generated by a femtosecond laser system is split into two paths by a polarizing beam splitter (PBS) and independent intensity control in each path is achieved with half-wave plates ($\lambda/2$). An optical delay line (DL) adjusts the mutual time delay ($\Delta t$) between laser pulses in the paths, which are spatially recombined by a second PBS and focused onto the sample (S) using a lens (L). **b**, Electrical readout setup for stroboscopic experiments. The sample's resistance $R_{sample}$ is inferred from oscilloscope waveform measurements of the voltage $U$ across the input resistance $R_{scope}$ = 50 Ω. During measurements, a voltage source with $U_{source}$ = 150 mV is utilized.

## Note S1. Switching threshold model and effect of local resistivity change on device resistance

### S1.1. Switching threshold model

In Fig. S2a, we assume a Gaussian lateral profile for the incident laser pulse, with a half-width $w \approx 10$ μm. Due to the absorption in CuMnAs epilayer, the pulse intensity decays exponentially with an absorption coefficient of $\alpha = 3.1 \cdot 10^5$ cm$^{-1}$ [S1]. The resulting energy distribution of the laser light (expressed as energy density per unit volume) within the epilayer is described by

$$U(r,z) = U_0 \cdot e^{-\frac{r^2}{w^2}} \cdot e^{-\alpha z}. \qquad \text{(S1.1)}$$

Here, $r$ is the radial distance from the center of the Gaussian profile, $z$ denotes distance in the epilayer along the laser propagation direction, and $U_0$ is the amplitude of the energy density.

To evaluate the fluence dependence of the resistive signal (Fig. 2b), we assume that the signal is proportional to the volume of the region in which the energy density in CuMnAs surpasses a threshold value, $U(r, z) \geq U_{TH}$. This volume is given by

$$V_{\text{switched}} = \frac{\pi w^2}{2\alpha} \ln^2(\theta)\,;\ \theta \geq 1, \qquad \text{(S1.2)}$$

where $\theta$, defined as $\theta = U_0/U_{TH}$, measures the extent to which the threshold energy is exceeded. This quantity can also be represented in terms of the incident laser fluence $F$ (energy density per unit area) as $\theta \approx F/F_{TH}$, where $F_{TH}$ is the threshold fluence. If $\theta < 1$, the threshold condition for achieving the switching is not fulfilled, implying $V_{\text{switched}} = 0$.

This switched volume can be envisaged as a cap-like shape (see Fig. S2) with a circular base of diameter $d$:

$$d = 2w\sqrt{\ln(\theta)};\quad \theta \geq 1. \qquad \text{(S1.3)}$$

Fig. S2b demonstrates how the diameter $d$ expands as the switching threshold is exceeded. Despite the relatively large spot size of the incident laser pulse, a considerably smaller region can be switched.

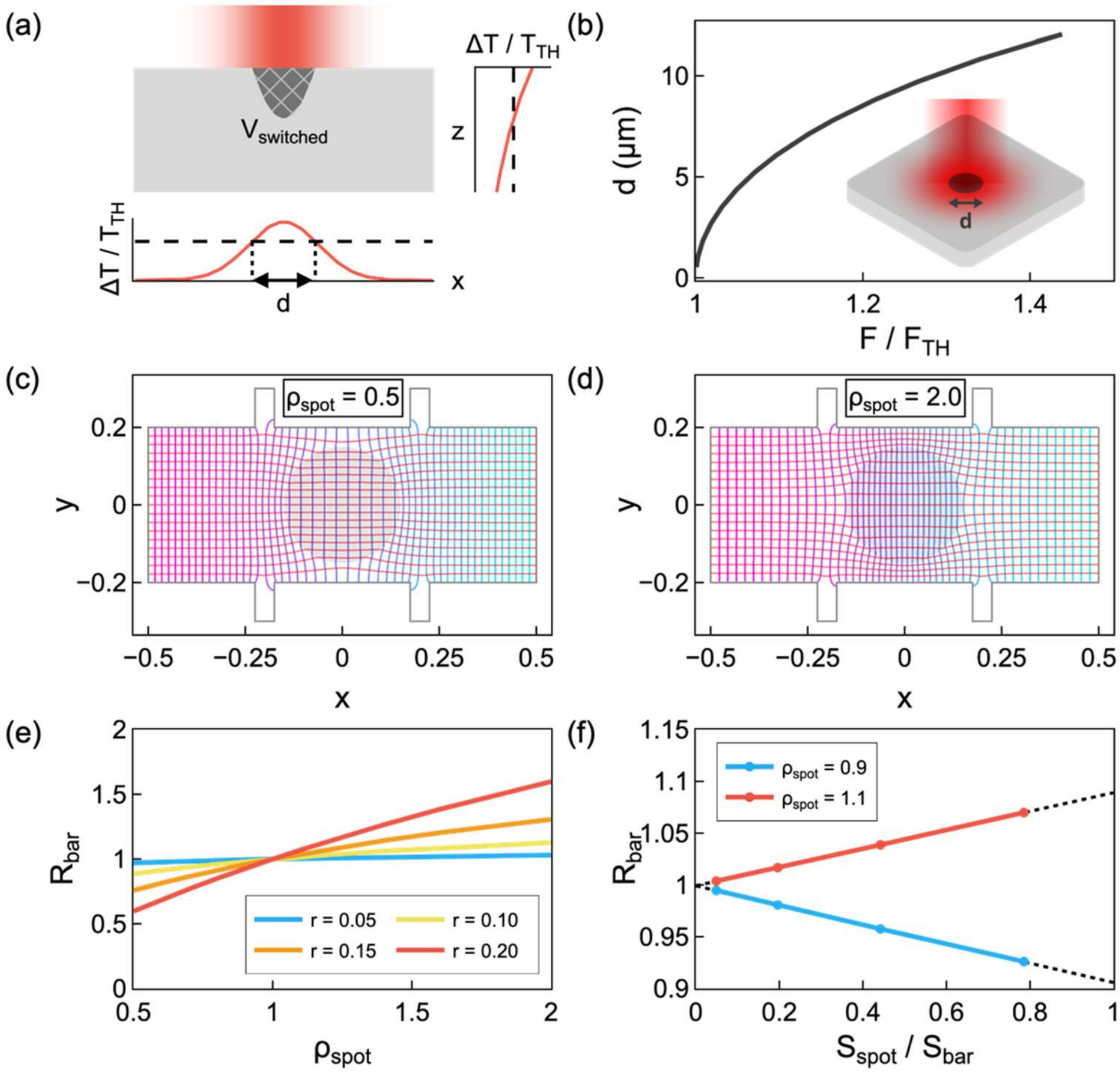

**Fig. S2 | Linear proportionality between the measured resistive signal and the volume of area with changed resistance. a,** The incident laser pulse with a Gaussian lateral ($x$) profile is absorbed in the CuMnAs epilayer with an exponentially decaying intensity profile in the $z$-direction. The volume $V_{\mathrm{switched}}$ represents the region where the laser-pulse-induced temperature increase $\Delta T$ of CuMnAs exceeds the threshold temperature $T_{\mathrm{TH}}$. This volume can be visualized as a cap-like structure with a circular base of diameter $d$. **b,** The dependence of $d$ on fluence $F$, expressed relative to the value of the threshold fluence $F_{\mathrm{TH}}$. Note that despite a relatively large spot size of the incident laser pulse ($w$ = 10 μm), a considerably smaller region of CuMnAs can be switched. **c,** Examples of potential distribution and the corresponding current lines in a device structure with a circular spot in the middle (with a radius $r = 0.15$), where the local resistivity $\rho$ is decreased from $\rho = 1.0$ to $\rho = 0.5$ or **d,** increased from $\rho = 1.0$ to $\rho = 2.0$. **e,** For a fixed spot radius $r$, the calculated apparent resistance $R_{\mathrm{bar}}$ of the device scales monotonously with the resistivity inside the spot $\rho_{\mathrm{spot}}$. **f,** For a fixed value of $\rho_{\mathrm{spot}}$, which deviates weakly from that in the rest of the device, the apparent resistance scales linearly with the spot area $S_{\mathrm{spot}} = \pi r^2$.

### S1.2. Estimation of threshold switching energy

Determining the switching threshold fluence $F_{TH}$ enables us to estimate the minimal excitation energy required to induce the switching in CuMnAs. Fluence threshold $F_{TH}$ = 18.5 mJ/cm$^2$ (evaluated further using method in Note S2) corresponds to ≈ 60 nJ energy per pulse incident on the surface of the CuMnAs wafer. Taking into an account the reflectance of the wafer $R \approx 40$ % (see Fig. 6c in Ref. S1), the total energy delivered by the single laser pulse at switching threshold $\epsilon_{TH}$ is ≈ 36 nJ. Considering the energy distribution of the laser light $U(r, z)$ within the epilayer described by Eq. (S1.1), the total delivered energy $\epsilon_{TH}$ can be expressed as

$$\epsilon_{TH} = \int_V U(r,z)\, dV = \frac{\pi w^2}{\alpha} U_{TH}, \tag{S1.4}$$

where $U_{TH}$ is the amplitude of the energy density at switching threshold. Consequently, the threshold energy density can be evaluated as $U_{TH} \approx 3.5$ kJ/cm$^3$.

### S1.3. Effect of local resistivity change on device resistance

The assumption, mentioned in Note S1.1, that the resistive signal measured across the device is linearly proportional to the volume of the switched area with altered resistance, is not self-evident. To validate this, we conducted simulations modelling a similar situation. Considering the relative thickness of the utilized epilayers (~ nm) compared to the lateral dimensions of the device (~ µm), simulations were executed in 2D, neglecting the effect of depth. Potential and current distributions were calculated using the finite-element method in a bar-shaped device structure with current contacts at its left and right edges, and with a pair of thin vertical arms to probe the potential drop along the sample, see Figs. S2c and S2d. This potential drop and the calculated total current were utilized to evaluate the four-point (apparent) resistance of the device. The true resistivity of the device material was assumed to be non-uniform: equal to 1.0 everywhere except within a circular spot in the middle of the device, where the resistivity was locally increased (Fig. S2c) or decreased (Fig. S2d). The calculated dependences of the apparent resistance $R_{bar}$ on the spot resistivity and spot size are shown in Figs. S2e and S2f, respectively. It indicates that when the change in the spot resistivity $\rho$ is relatively small (here plus or minus 10% of the value in the unperturbed material), the device resistance scales linearly with the spot area, $S_{spot} = \pi r^2$, and approaches the spot resistivity if the spot covers the whole area between the potential probes $S_{bar}$.

**Note S2. Single-pulse method: Separation of switching and heat contributions from measured resistance change**

Fig. 1c presents as-measured oscilloscope data for three selected fluence values of the incident single laser pulse in a device fabricated from a 20-nm-thick CuMnAs epilayer. The major resistive signal following the incidence of the laser pulse can be attributed to the light-induced increase in the sample's temperature. This excess heat dissipates entirely in ≈ 1 μs, as apparent in the data for a fluence of 20 mJ/cm$^2$ in Fig. 1c. If a specific laser pulse fluence threshold is exceeded, a new long-lived resistive signal appears, which persists even in the time region where the excess heat is fully dissipated – see data for fluences 30 mJ/cm$^2$ and 40 mJ/cm$^2$ in Fig. 1c. We ascribe this persistent component to the switching signal. As evident from Fig. 1c, the heat and switching signals are intermixed for times < 1 μs, and the challenge lies in separating their contributions to the measured resistive signals.

Fig. S3a presents the same data set as in Fig. 1c, but it is plotted as a function of fluence $F$ for three selected times $t$ after the incidence of the laser pulse. The data exhibit a characteristic onset with a fluence approximately at 20 mJ/cm$^2$. In addition, there is also a linear background, particularly apparent for short times $t$ after the incidence of the laser pulse (black lines in Fig. S3a). This background can be attributed to the device resistance change due to temperature increase, which can be approximated by a linear function from 300 K to ≈ 400 K (see Fig. S4a). The energy delivered to CuMnAs by the laser pulse is characterized by its fluence and, therefore, the heat contribution to the resistance change can be approximated by

$$\Delta R_{(\mathrm{heat})} \sim k * F. \qquad \text{(S2.1)}$$

Here, $k$ represents the slope of the linear background. Fig. S3b displays these linear backgrounds for various times $t$, fitted for fluences up to 15 mJ/cm$^2$. The inset of Fig. S3b depicts the time evolution of the slope $k$, which represents a measure of the excess heat in the studied device. This indicates that the heat is fully dissipated within ≈ 1 μs after the impact of the laser pulse.

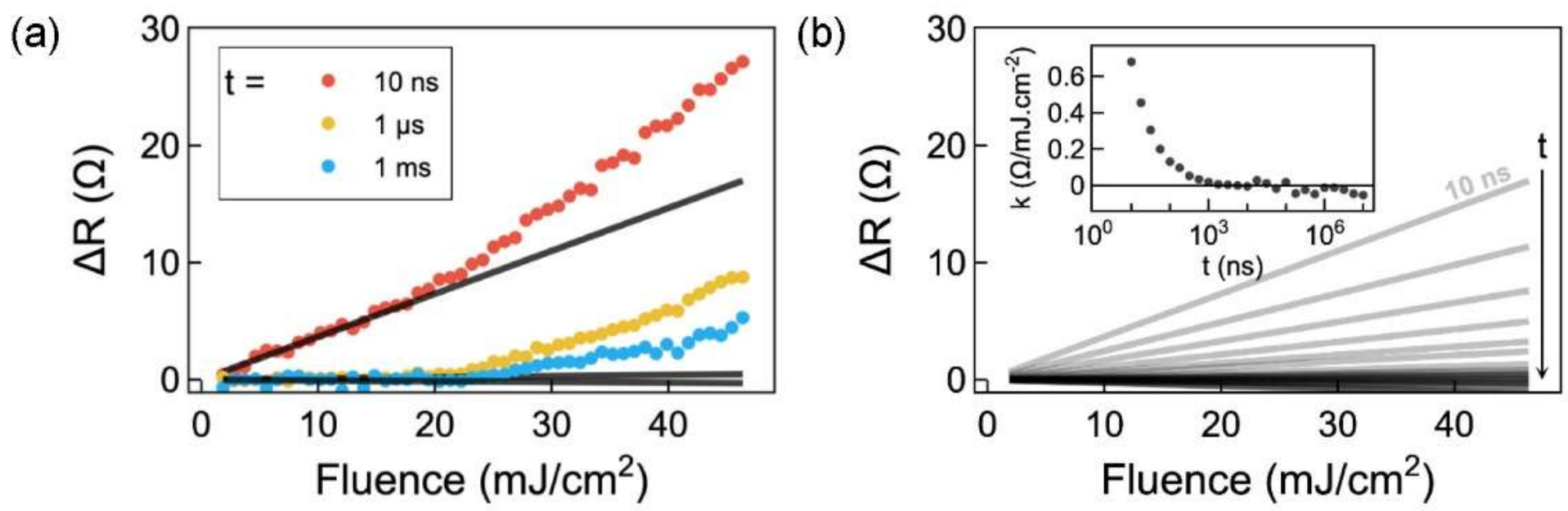

**Fig. S3 | Separation of the heat-related resistive signal appearing as a linear background in the as-measured data. a,** The as-measured fluence dependence of resistive signal change $\Delta R$ at three different times $t$ after the incidence of a single laser pulse (dots) in a device fabricated from a 20-nm-thick CuMnAs epilayer. The linear backgrounds (lines) are attributed to the heat contributions to the resistive signal. These backgrounds were determined by fitting the data up to a fluence of 15 mJ/cm$^2$ using Eq. (S2.1). **b,** The linear backgrounds at various times $t$. Inset: The time evolution of the slope $k$ of the linear background, representing the measure of excess heat in the investigated device. The heat is fully dissipated within ≈ 1 μs after the impact of the laser pulse.

The switching contribution to the device resistance can be determined by subtracting the linear background from the as-measured data, as shown in Fig. 2b. By re-plotting the data as a function of time $t$, this method allows us to deduce the time evolution of heat and switching separately, as depicted in Figs. 2c and 2d. The relaxation of the switching signal can be described by the Kohlrausch stretched exponential [S2]:

$$\Delta R_{(\mathrm{switching})}(t) \sim A_{S1} e^{-\left(\frac{t}{\tau_{S1}}\right)^{0.6}} + A_{S2} e^{-\left(\frac{t}{\tau_{S2}}\right)^{0.6}} \quad \text{(S2.2)}$$

with two dominant components characterized by different relaxation times $\tau_{\mathrm{S1}}$, $\tau_{\mathrm{S2}}$ and corresponding weights $A_{\mathrm{S1}}$, $A_{\mathrm{S2}}$. These relaxation times are temperature dependent, with slow and fast room-temperature relaxation times in the approximate ranges of ~ 10 s and ~ 10 ms, respectively (see Fig. 4 in Ref. S2). The relaxation of the switching signal in Fig. 2d is described by the fast component with a relaxation time of ≈ 4.5 ms, whereas the slow component is treated as a constant within the measured 10 ms time window. As shown in Ref. S2, the temperature dependence of relaxation times of both of the components can be analyzed in terms of thermally activated relaxation processes, and the obtained attempt rate suggests a connection to the ultrafast antiferromagnetic dynamics. The reasons for a presence of two distinct components in the relaxation process remain unknown. Possibly, it could be related to the presence of more types of spin textures, such as domain walls oriented along different crystallographic planes and/or their diffusion in different directions. Another possibility is the role of characteristic crystallographic defects such as Cu/Mn swaps or dislocations active as centers for domain wall pinning and/or domain wall annihilation. A comprehensive understanding of the physical origin of the components necessitates further advancements in magnetic imaging techniques with atomic resolution [S3].

Although the excess heat in the sample apparently scales with the energy delivered by the light pulse, the linear scaling of the resulting resistive signal according to Eq. (S2.1) is not obvious. Fig. S4a displays the temperature dependence of CuMnAs resistance, measured in a separate electrical experiment during which the sample underwent a gradual heating process. Here, the Néel temperature $T_N \approx 453$ K was estimated by evaluating the inflection point of this dependence. The measured resistance change is linear in temperature approximately up to ≈ 400 K (as shown by the black line). The inset of Fig. S4a presents the data after subtracting the linear function to emphasize the existence of a non-linear component in the measured dependence. Consequently, the linear scaling of the resistive signal following Eq. (S2.1) applies only if temperature fluctuations within the sample remain relatively low (i.e., when the temperature increase from 300 K remains below ≈ 400 K). Fig. S4b presents the fluence dependence of resistive signal at the moment of the laser pulse impact (i.e., at the peak of the resistive signal change observed at $t = 0$ ns; see also Fig. S5). The depicted value of the fluence threshold $F_{TH}$ was determined using the switching threshold model (see Fig. 2b and Note S1). As clearly apparent from this figure, even a laser pulse with a fluence below the switching threshold can significantly elevate the sample's temperature at the moment of incidence, leading to a non-linear temperature dependence of the measured resistance. The non-linear region can be again identified by subtracting a linear function from the measured dependence, which is shown in the inset of Fig. S4b. Overall, the similarities between insets in Fig. S4a and Fig. S4b suggest that there exists a relationship between the fluence threshold $F_{TH}$ and the Néel temperature $T_N$.

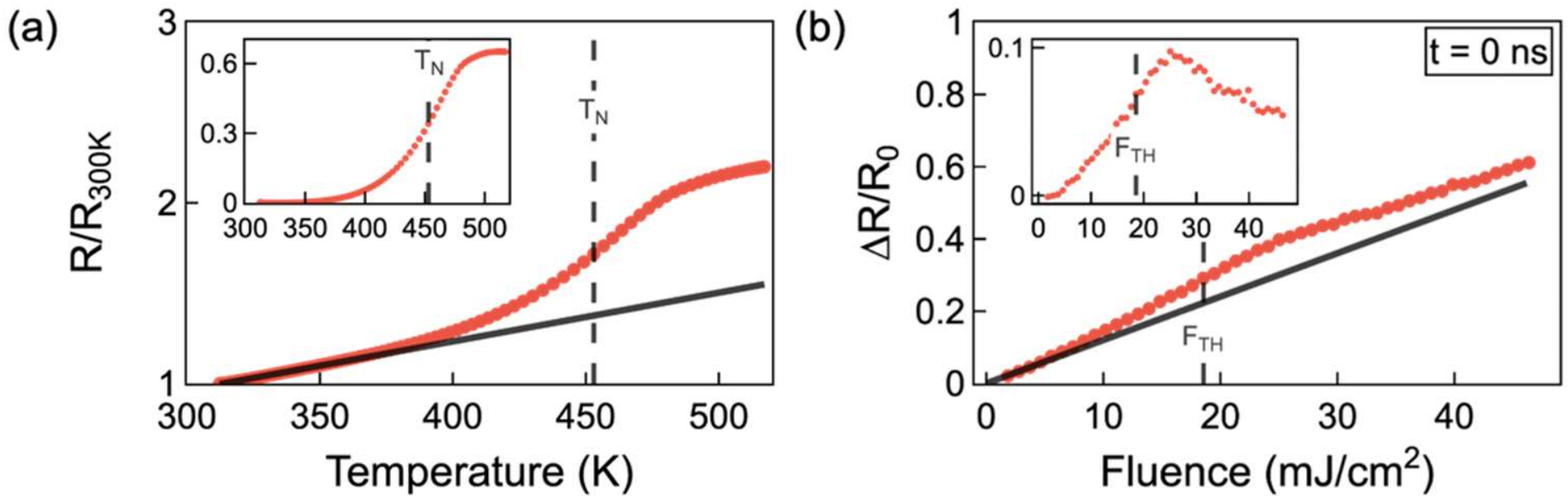

**Fig. S4 | Dependence of the heat-related resistive signal on sample temperature and delivered laser pulse energy. a,** The temperature dependence of CuMnAs resistance (dots), normalized to the room-temperature value, as observed in an independent electrical experiment in which the sample was gradually heated. The Néel temperature $T_N \approx 453$ K was deduced by determining the inflection point in the obtained dependence. The black line represents the linear dependence in the temperature region up to ≈ 400 K. Inset: The non-linear component of the relative resistance change obtained by subtracting the linear background from the measured data depicted in the main figure. **b,** The fluence dependence of the resistive signal $\Delta R$, normalized to the base resistance $R_0$, at the moment of the laser pulse impact ($t = 0$ ns). The fluence threshold $F_{TH}$ is evaluated using the switching threshold model (see Note S1). Inset: The non-linear component of the signal is obtained by subtracting the linear background (line), fitted for fluences up to 5 mJ/cm$^2$, from the data.

Finally, we would like to comment on the overall time resolution of this procedure, which enables the separation of switching and heat contributions from the as-measured resistances. To be applicable, the linear relationship between the laser-pulse-induced sample heating and the resulting resistance change, as described by Eq. (S2.1), has to be fulfilled. In a 20-nm-thick CuMnAs epilayer, this is the case for times $t \geq 10$ ns, when the vast majority of heat deposited to the sample is already dissipated (see Fig. 2c) and, presumably, the temperature of the excited spot is reduced below ≈ 400 K (see Fig. S4a). Technically, this technique can be applied also for shorter times but the reliability of the obtained outputs is questionable. Therefore, we depicted this unreliable time range by a grey color in Figs. 2c and 2d.

**Note S3. Double-pulse method: Heat accumulation and separation of resistive signals for two mutually-time-delayed laser pulses**

The CuMnAs device structure can serve as a memory element, capable of encoding analog information using a pair of femtosecond laser pulses mutually-time-delayed on ultrafast timescales (see Fig. 1e). The writing process in these devices is governed by the ultrafast heat dynamics in CuMnAs, as schematically illustrated in the inset of Fig. 1e. Depending on the pulse-pair mutual time-delay, the accumulated heat in CuMnAs can lead to a temperature increase above a certain threshold temperature $T_{\mathrm{TH}}$ and, consequently, to the switching. In this Chapter we explore this effect experimentally. Moreover, we illustrate how this two-pulse experiment can be utilized to deduce switching related resistivity dynamics, which is in a very good agreement with the results of a single-pulse method depicted in Fig. 2 and analyzed in Note S2.

As extensively discussed in Note S2, the absorption of a femtosecond laser pulse leads to a transient temperature increase in the CuMnAs epilayer. This temperature alternation consequently results in a time-dependent resistivity change (cf. Fig. S4) detected in our experiment (see Fig. S5). Due to the limited bandwidth of the electrical setup, the ultrafast (picosecond) local heating of the material (see Figs. 3c and 3d) after laser pulse impact is apparent in the measured resistivity waveforms as a peak with a ≈ 2 ns-long rising edge, representing the time resolution of our electrical experiment depicted as a gray area in Fig. S5a. (The sub-nanosecond heat dynamics is discussed in detail in Note S4.) Up to ≈ 10 ns, the laser pulse-induced resistivity change is dominated by the heat contribution (cf. Figs. 2c and 2d) originating in sample's increased temperature. As the resistivity is a monotonous function of the sample temperature (see Fig. S4a), the early stages of the as-measured resistivity dynamics depicted in Figs. S5a and S5b can be regarded, to some extent, as a visualization of the temperature dynamics in the sample. In particular, the "summation" of heats delivered by a pair of laser pulses, mutually delayed by a time $\Delta t$, can be studied experimentally directly in a time domain. Here, we use a notation where laboratory time $t = 0$ is defined by the position of the resistivity peak induced by the impact of the second laser pulse, which is denoted as $B$ in Fig. S5. If the first laser pulse, denoted as $A$, is incident on the sample before pulse $B$, we consider their mutual time delay $\Delta t > 0$. In this manner we can deduce the evolution of their *combined* effect in a laboratory time $t$, as depicted in Figs. S5 and S6, and discussed in detail below.

Experimentally, the resistivity changes induced by two identical $\Delta t$-time-delayed laser pulses can be recorded either for the separate impact of each pulse (waveforms $A$ and $B$ in Figs. S5a and S5b) or when both pulses are present (waveform $AB$). Waveform $A+B$ represents the arithmetic sum of waveforms $A$ and $B$. Fig. S5a demonstrates the heat accumulation following the incidence of two laser pulses with a large time-delay $\Delta t$ = 4 ns. In this case, the majority of the heat delivered by pulse $A$ dissipates before the incidence of pulse $B$. As a result, the cumulative heat does not surpass the threshold level (schematically depicted as a dashed horizontal line) necessary to switch the sample. Consequently, the arithmetic sum of individual heat contributions from pulses $A$ and $B$ closely matches the measured waveform $AB$. In contrast, for a small time-delay between pulses ($\Delta t$ = 10 ps in Fig. S5b), the cumulative heat exceeds the threshold, and the arithmetic sum of the waveforms $A$ and $B$ deviates considerably from the waveform $AB$ measured when both pulses are present. Despite the fact that neither pulse $A$ nor pulse $B$ can trigger switching individually, the combined effect of this pulse-pair can. Hence, the difference between waveforms $AB$ and $A+B$ corresponds to the switching component in the resistive signal

$$\Delta R_{\mathrm{S}}(t) \sim \Delta R_{\mathrm{AB}}(t) - [\, \Delta R_{\mathrm{A}}(t) + \Delta R_{\mathrm{B}}(t)]. \qquad \text{(S3.1)}$$

The applicability of this approach, which is based on the assumption that the resistivity is approximately linearly dependent on the delivered heat (i.e., on the increase in sample temperature), is illustrated in Fig. S5c. Here we compare results from this double-pulse method with those of the single-pulse method described in Note S2. Clearly, the dynamics of the switching signals derived from both methods show a very good agreement. Note that in the case of the double-pulse method, the total double-pulse fluence $F^*$ (defined as a sum of the fluences of both pulses) exceeds the single-pulse fluence $F$ necessary to achieve comparable resistive signals (e.g., 46 vs. 40 mJ/cm$^2$ for $\Delta t$ = 10 ps in Fig. S5c). This difference in total delivered energy originates from the partial dissipation of heat from the first pulse before the incidence of the second. (Due to the ratio of the ≈ 10 μm excitation spot diameter and the tens of nm thicknesses of the CuMnAs films, we consider the one-dimensional heat diffusion from the CuMnAs film to the GaP substrate to dominate this – see the heat transfer simulations in Ref. S1).

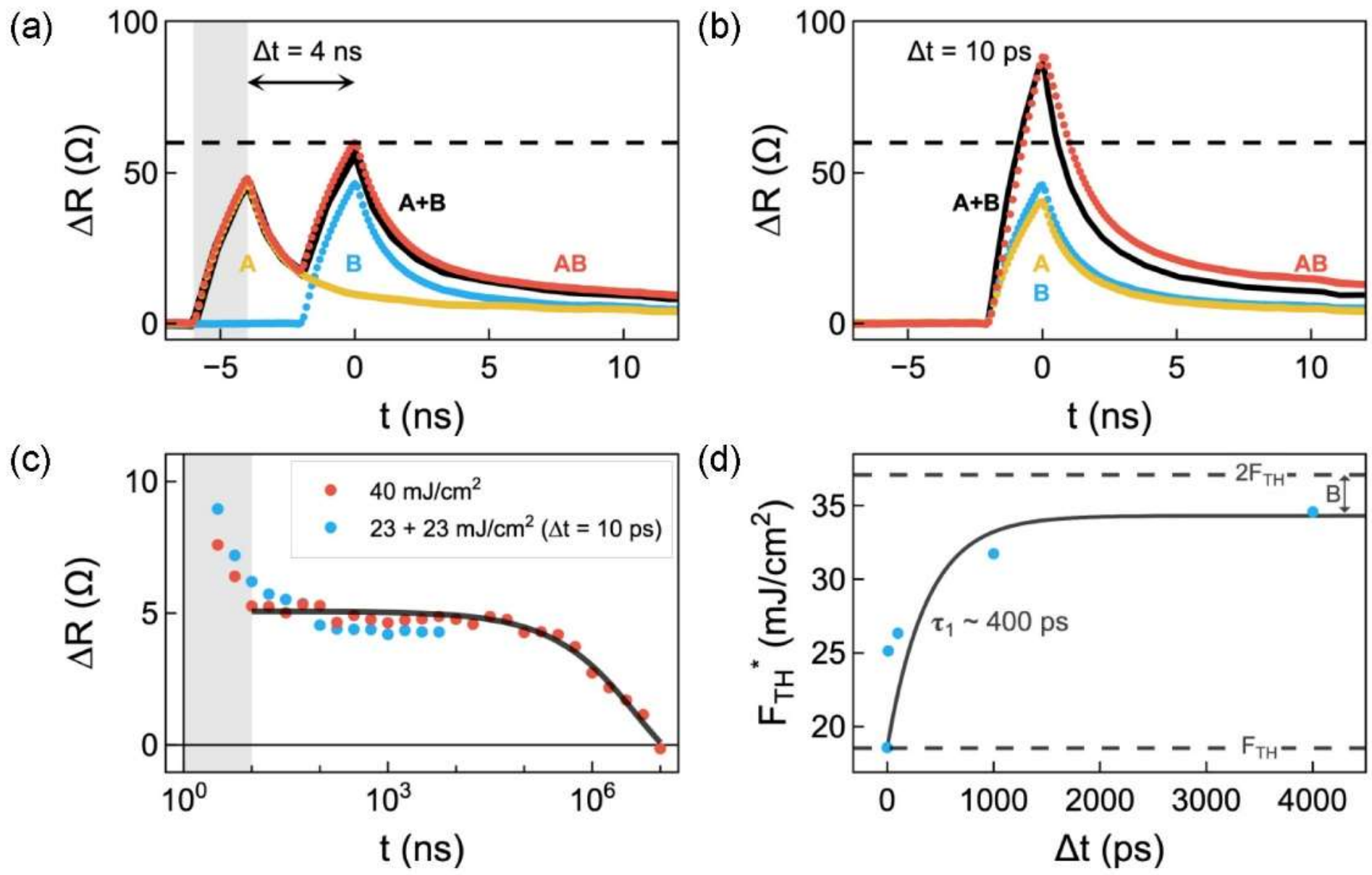

**Fig. S5 | Accumulation of heat from two identical, mutually-time-delayed laser pulses in a device prepared from a 20-nm-thick CuMnAs epilayer. a,** Time-resolved resistive signals measured for two identical laser pulses with fluence 20 mJ/cm$^2$ and a mutual time-delay $\Delta t$ = 4 ns. The waveforms $A$ and $B$ depict the effect of each pulse's incidence individually, while the waveform $AB$ represents the resistivity change when both pulses are present. The waveform $A+B$ represents the arithmetic sum of waveforms $A$ and $B$. The dashed horizontal line illustrates the resistive threshold level which needs to be surpassed to trigger the measurable switching in the sample. The electrical measurements' time resolution of ≈ 2 ns is depicted as a grey area (see text). **b,** Same as **a**, but for $\Delta t$ = 10 ps. Again, the laser pulses are unable to trigger the detectable switching individually. However, their combined effect exceeds the threshold $\Delta R_{TH}$ (waveform $AB$). The difference between waveforms $AB$ and $A+B$ is due to the resistivity change induced by the switching, which is depicted in **c** as green points. Red points in **c** represent resistivity dynamics deduced from the single-pulse method described in Note S2. The gray area denotes the time window where the applicability of the used methods is questionable (see Note S2). The black line describes the relaxation of the switching signal using the Kohlrausch stretched-exponential model (see Note S2). **d,** The total delivered fluence threshold $F_{TH}^*$ for a pair of laser pulses shown as a function of their mutual time delay (points); see Fig. S6 for the measured data from which the values of $F_{TH}^*$ were deduced. The threshold ranges from the single-pulse threshold value $F_{TH}$, for time-overlapping pulses, to $2F_{TH}$ for very distant pulses. The line represents the theoretical model of Eq. (S3.3) with a thermal relaxation time constant $\tau_1$ ~ 400 ps, see text.

The deduced dynamics of switching signals for three values of $\Delta t$ are displayed in Figs. S6a, S6c and S6e. The corresponding fluence dependencies, depicted in Figs. S6b, S6d, and S6f, reveal the aforementioned inter-pulse partial heat dissipation effect, which is apparent as an increase of the total delivered fluence threshold $F_{TH}^*$ with $\Delta t$. This dependence is summarized in Fig. S5d as points. Intuitively, the plausible extreme values of $F_{TH}^*$ (represented as dashed horizontal lines in Fig. S5d) can be understood as follows. If $\Delta t$ approaches 0 ps, the situation evolves towards a single laser pulse scenario, where $F_{TH}$ is sufficient to switch the sample, consequently leading $F_{TH}^*$ to converge to $F_{TH}$. Conversely, for $\Delta t$ considerably exceeding the heat dissipation times, the effect induced by both pulses become entirely independent. Now, to achieve a switching, each pulse has to exceed the threshold $F_{TH}$ independently, thereby causing $F_{TH}^*$ to converge towards $2F_{TH}$. To derive the expected dependence between these two extreme values, we need to estimate the energy, delivered by the first pulse, that partially dissipates before the impact of the second pulse at time delay $\Delta t$. This can be expressed as

$$\Delta E(\Delta t) = A - \left( (A - B) e^{-\frac{\Delta t}{\tau_1}} + B \right) \quad \text{(S3.2)}$$

Here, $A$ is the amplitude of the heat energy delivered by the pulse, and $B$ is the residual heat background (see Fig. S9a and Note S4 for a detailed description). If the first pulse is delivering the threshold fluence $F_{TH}$, the energy dissipated after $\Delta t$, which the second pulse must compensate for, can be expressed as $\Delta E_{TH}(\Delta t) = \Delta E(\Delta t \mid A{=}F_{TH}, B{=}B_{TH})$. The total fluence threshold is then:

$$F_{TH}^*(\Delta t) = F_{TH} + \Delta E_{TH}(\Delta t) = F_{TH} + (F_{TH} - B_{TH}) * \left(1 - e^{-\frac{\Delta t}{\tau_1}}\right) \quad \text{(S3.3)}$$

The model described by Eq. (S3.3) was utilized to generate the solid line in Fig. S5d using values $F_{TH} = 18.5$ mJ/cm$^2$ (obtained from the switching threshold model described in Note S2), $\tau_1 = 375$ ps (inferred from the heat dissipation model in Note S4), and a fitted value of $B_{TH} = 2.8$ mJ/cm$^2$.

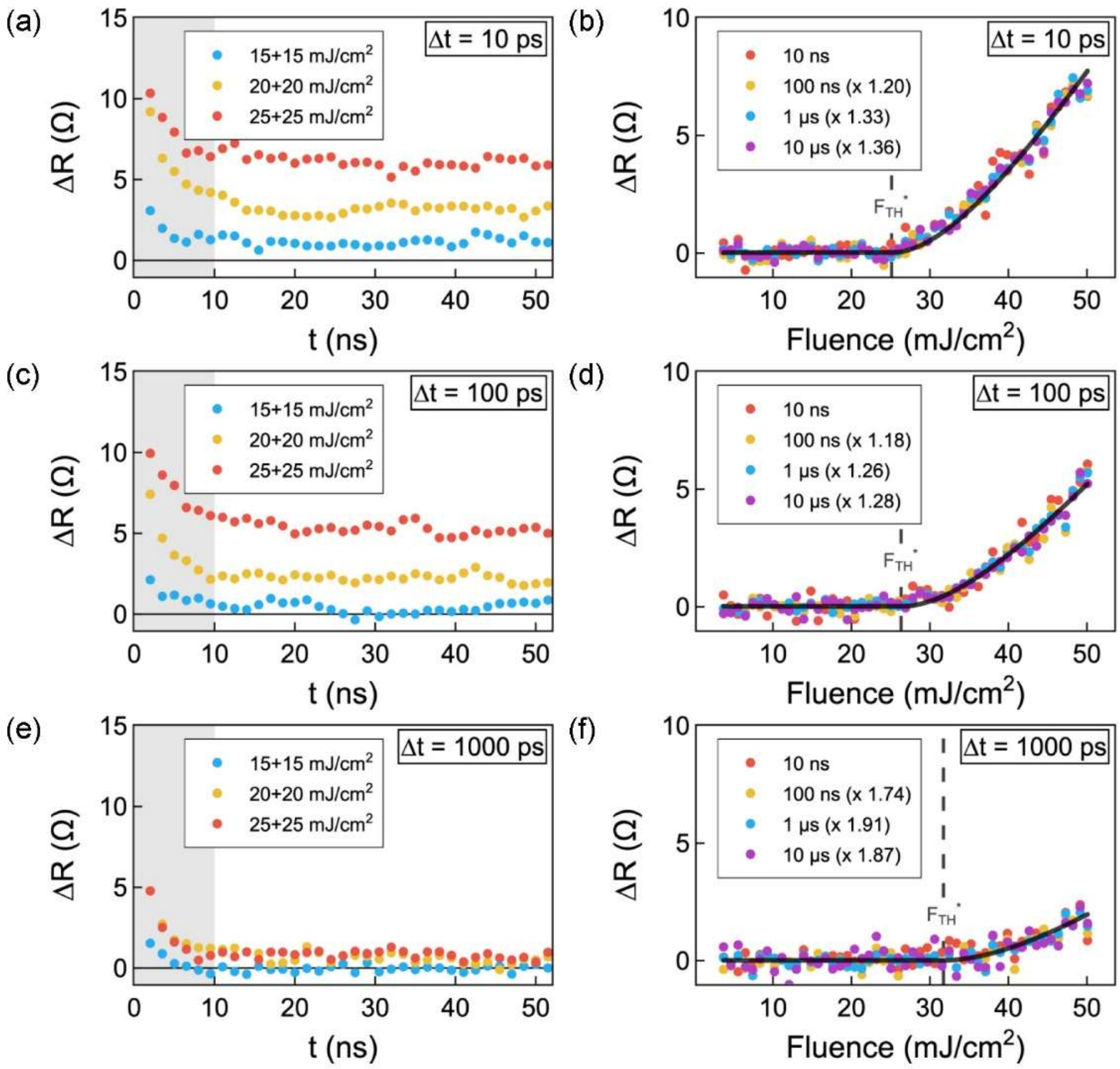

**Fig. S6 | Switching signals in a device prepared from a 20-nm-thick CuMnAs epilayer sample measured by the double-pulse method. a,** Dynamics of switching signals obtained for a time delay between pulses $\Delta t$ = 10 ps and depicted fluences. **b,** Switching contributions of resistive signals, corresponding to the same $\Delta t$ as in **a,** shown as a function of the total fluence from both pulses for depicted times $t$ after the impact of the second pulse. The data are multiplied by the indicated factors to show their common fluence dependence. The line represents a fit using a theoretical threshold model (see Note S1), defining the value of the total fluence threshold $F_{TH}^*$. **c, d,** and **e, f,** are analogous to **a** and **b** with $\Delta t$ = 100 ps and 1000 ps, respectively.

In conclusion, this Chapter aimed to illustrate that a rather complex experimental data measured using a pair of mutually-time-delayed laser pulses (Figs. S5 and S6) can be modeled well by a simple phenomenological model describing the heat dissipation dynamics in CuMnAs. The consequences of this are discussed in detail in Note S4.

**Note S4. Heat dissipation dynamics**

Optical writing experiments employing a pair of laser pulses, as depicted in Fig. 3, reveal that the switching in CuMnAs epilayers is governed by ultrafast heat dynamics: incident optical pulses always heat the sample, but they trigger the writing only when a specific excitation threshold is surpassed (cf. inset of Fig. 1e). However, deducing the heat dynamics directly from these measurements is not straightforward. The complexity arises because $\Delta R$ signal is influenced by the nonlinear onset response (see Fig. 2b), making it highly sensitive to exceeding the threshold. A clear manifestation of this enhanced sensitivity can be observed, for example, in a device prepared from a 20-nm-thick CuMnAs epilayer, where the half-width at full maximum of the $\Delta R$ dependence on $\Delta t$ is approximately 90 ps (see Fig. 3a), in contrast to the thermal dissipation time $\tau_1 \approx 400$ ps, measured in an independent experiment depicted in Fig. 3d.

**S4.1. Effective switching fluence in the double pulse experiment**

To address this, we introduce a concept of the effective switching fluence $F_{\mathrm{eff}}(\Delta t)$. This quantity represents the fluence of a single laser pulse that would generate the same $\Delta R$ signal as a laser pulse-pair with a given mutual-time-delay $\Delta t$.

Converting the resistive signal measured after the incidence of a laser pulse-pair $\Delta R_{\mathrm{2p}}(\Delta t)$ (double-pulse signal in the following) to $F_{\mathrm{eff}}(\Delta t)$ requires determining an analytical formula $F(\Delta R_{\mathrm{1p}})$, which is relating the fluence $F$ and the resistive signal measured after the incidence of a single laser pulse $\Delta R_{\mathrm{1p}}(F)$ (single-pulse signal in the following). Although the threshold model describing $\Delta R_{\mathrm{1p}}(F)$ (see Note S1) can, in principle, be utilized for this purpose, it does not guarantee the needed one-to-one invertible relationship between $F$ and $\Delta R_{\mathrm{1p}}$. To do so, we employed a sigmoidal function:

$$\Delta R_{\mathrm{1p}}(F) \sim \frac{A_{\mathrm{S}}}{1 + e^{-w_{\mathrm{S}}(F - F_{\mathrm{C}})}} \qquad \text{(S4.1)}$$

with a sigmoid amplitude $A_{\mathrm{S}}$, width $w_{\mathrm{S}}$ and center $F_{\mathrm{C}}$. Even though we do not ascribe any specific physical meaning to this model, it offers a good fit to the data. To determine the parameters of the sigmoidal model, we fitted $\Delta R_{\mathrm{1p}}(F)$ at multiple time points $t$ (see Fig. 2b) after the incidence of a single laser pulse, as illustrated in Figs. S7a, S7b, and S7c for a device prepared from a 20-nm-thick CuMnAs epilayer. Considering that apart from a scaling factor, the fluence dependencies at various times $t$ maintain the same shape (see Fig. 2b), we defined the sigmoid width $w_S$ and the center $F_C$ as global ($t$-independent) parameters, and the scaling factor $A_S(t)$ as a $t$-dependent parameter. Finally, we converted the double-pulse signal $\Delta R_{2p}(\Delta t)$ to the effective switching fluence $F_{eff}(\Delta t) = F(\Delta R_{2p}(\Delta t))$ using the $F(\Delta R_{1p})$ relation obtained by inverting the formula (S4.1), as shown in Fig. S7d. Note that this approach is applicable only in a region of non-zero $\Delta R_{2p}$ values and, hence the $x$-scale in Fig. S7d is considerably shorter than that in Fig. 3a. Fig. S8 illustrates the same process for a device prepared from a 50-nm-thick CuMnAs epilayer.

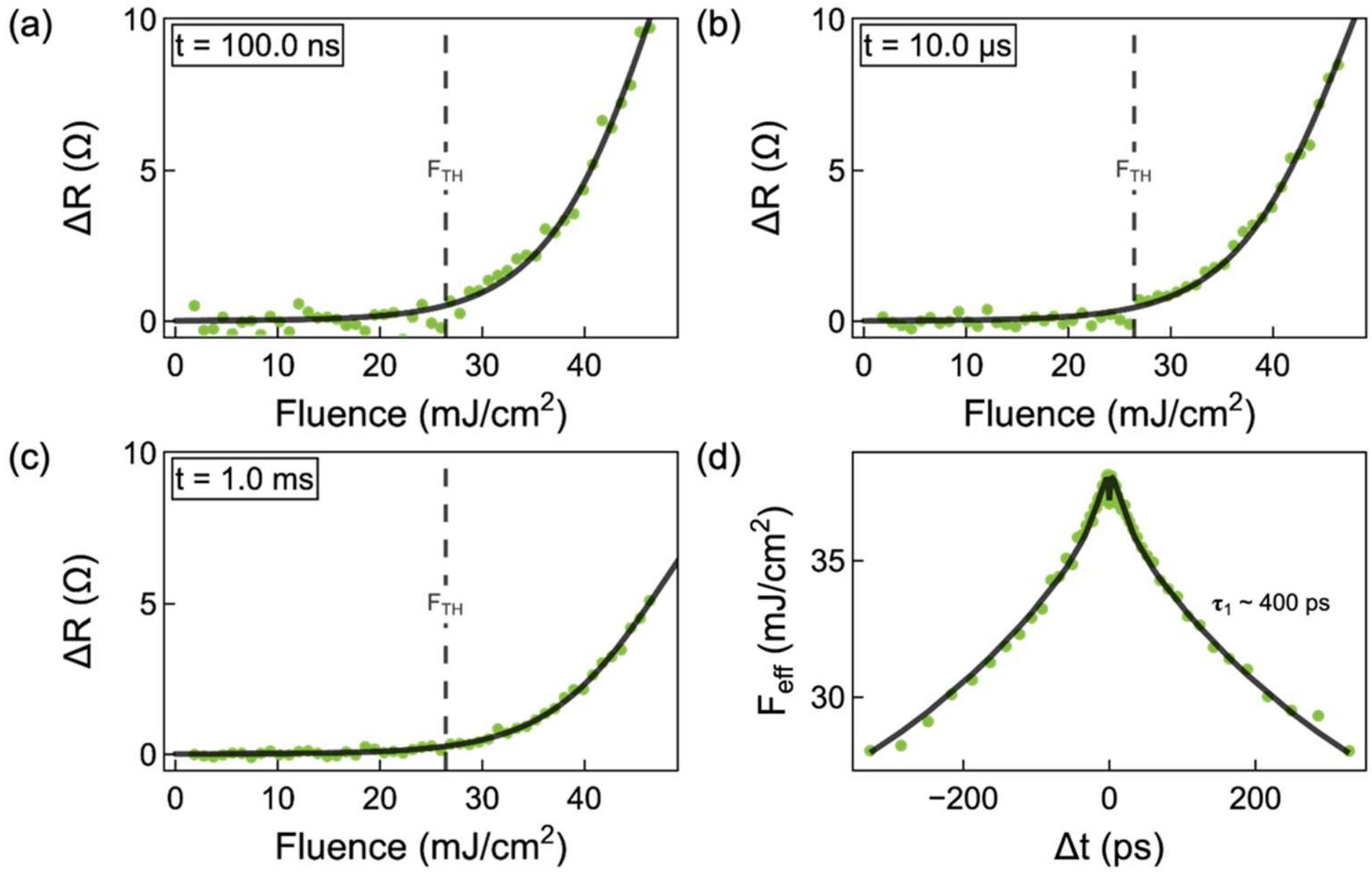

**Fig. S7 | Conversion of the measured double-pulse resistive signal to the effective switching fluence $F_{eff}$ in a device prepared from a 20-nm-thick CuMnAs epilayer. a - c,** Fluence dependence of $\Delta R$ at three times $t$ after the impact of a single laser pulse (dots). Data are fitted by Eq. (S4.1) to identify an analytical relationship between the fluence and $\Delta R$ (line). The obtained sigmoidal function with a width $w_s$ = 0.18 cm$^2$.mJ$^{-1}$ and a center $F_c$ = 47.7 mJ.cm$^{-2}$ was used to convert the double-pulse data into $F_{eff}$ using $A_S$ = 17.6 Ω for data at $t$ = 0.1 ms. The vertical line depicts the position of the fluence threshold $F_{TH}$. **d,** Obtained $F_{eff}$ as a function of time delay $\Delta t$ between the pair of laser pulses (dots). The line is a fit by Eq. (S4.5), yielding the depicted value of the heat dissipation time constant $\tau_1$.

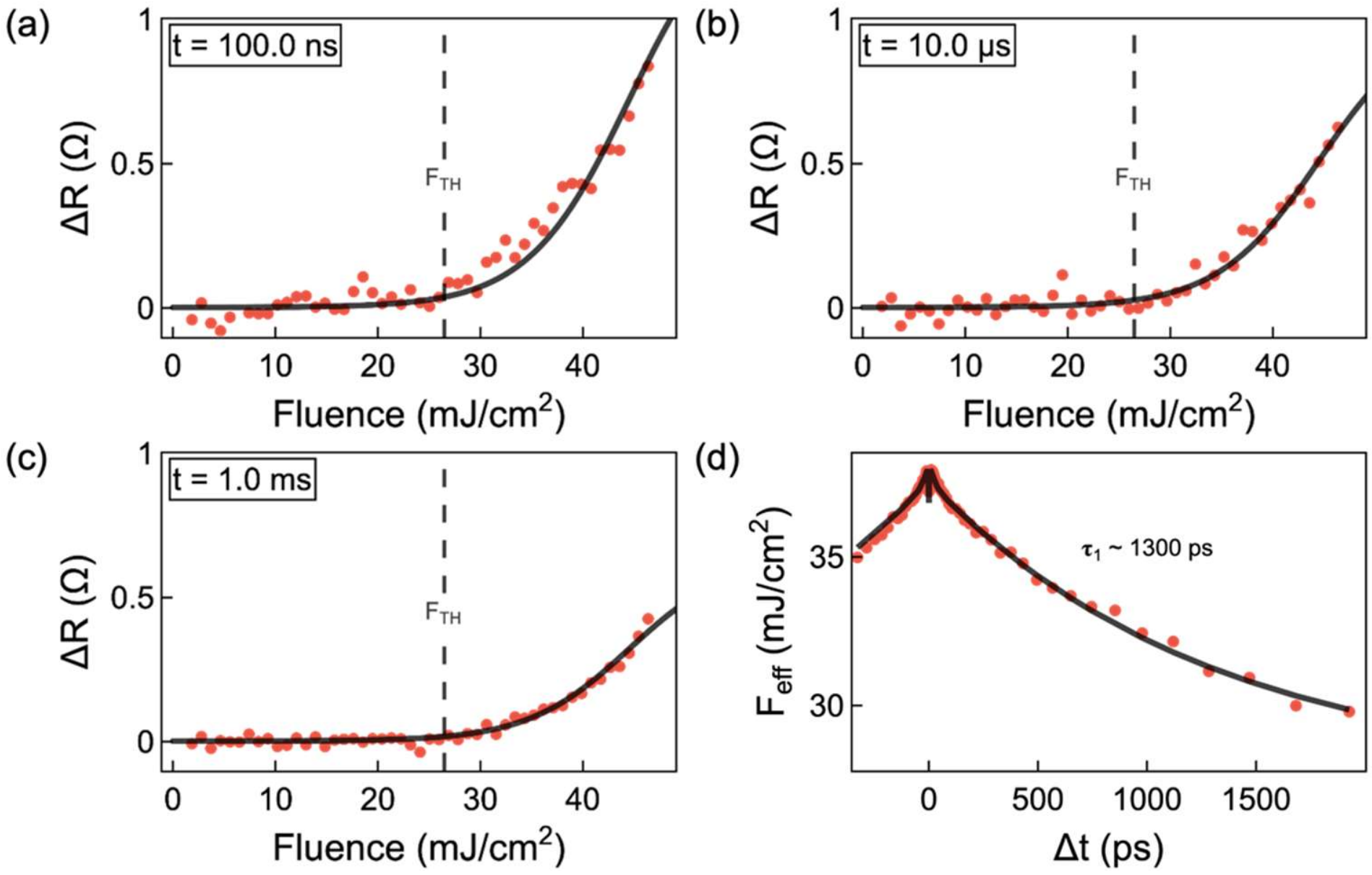

**Fig. S8 | Conversion of the measured double-pulse resistive signal to the effective switching fluence $F_{eff}$ in a device prepared from a 50-nm-thick CuMnAs epilayer. a - d,** Same as in Fig. S7 but for a 50-nm-thick CuMnAs epilayer; $A_S(0.1\text{ ms}) = 0.91\ \Omega$, $w_s = 0.20$ cm$^2$.mJ$^{-1}$ a $F_c = 44.5$ mJ.cm$^{-2}$.

### S4.2. Heat dynamics model

The effective switching fluence $F_{eff}$ in the double-pulse experiment estimates the effective energy delivered by the pulse-pair to the CuMnAs epilayer at the time of the arrival of the second pulse. Prior to the incidence of the second pulse, the heat originating from the first pulse partially dissipates. Assuming a monoexponential dissipation of the delivered heat from the epilayer to the substrate, characterized by a thermal relaxation time $\tau_1$, the relationship between $F_{eff}$ and the pulse-pair delay $\Delta t$ can be conceptually represented as:

$$F_{\text{eff}}(\Delta t) = F_1 e^{-\frac{\Delta t}{\tau_1}} + F_2, \tag{S4.2}$$

where $F_1$ and $F_2$ correspond to the incident fluences from the first and second laser pulses, respectively.

For our specific experimental conditions, we employed the phenomenological model depicted in Fig. S9a. We utilized identical laser pulses with fluences $F_1 = F_2 = A$, which are insufficient to trigger switching individually. However, when combined in a time overlap ($\Delta t = 0$), they result in an effective fluence of $2A$, well above the switching threshold. Due to the ratio of the ≈ 20 μm excitation spot diameter and the tens of nm thicknesses of the CuMnAs films, we are considering only one-dimensional heat diffusion from the epilayer. However, apart from the heat dissipation from the epilayer to the substrate, characterized by $\tau_1$ on ~ ns timescales (cf. Fig. 3d), we are also considering the heat dissipation from the substrate to the environment (e.g., to a sample holder and air), which occurs over longer timescales (see Fig. 2c). Therefore, on the ns timescales relevant for the double-pulse experiments, this residual heat background $B$ can be considered as a constant. Consequently, before the incidence of the second laser pulse, the heat originating from the first pulse dissipates as ~ $(A\text{-}B)\exp(-t/\tau_1)+B$ (see Fig. S9a), and the effective switching fluence $F_{\mathrm{eff}}(\Delta t)$ can be represented by the following phenomenological model:

$$F_{\mathrm{eff}}(\Delta t) \sim \left((A-B)e^{-\frac{\Delta t}{\tau_1}} + B\right) + A. \tag{S4.3}$$

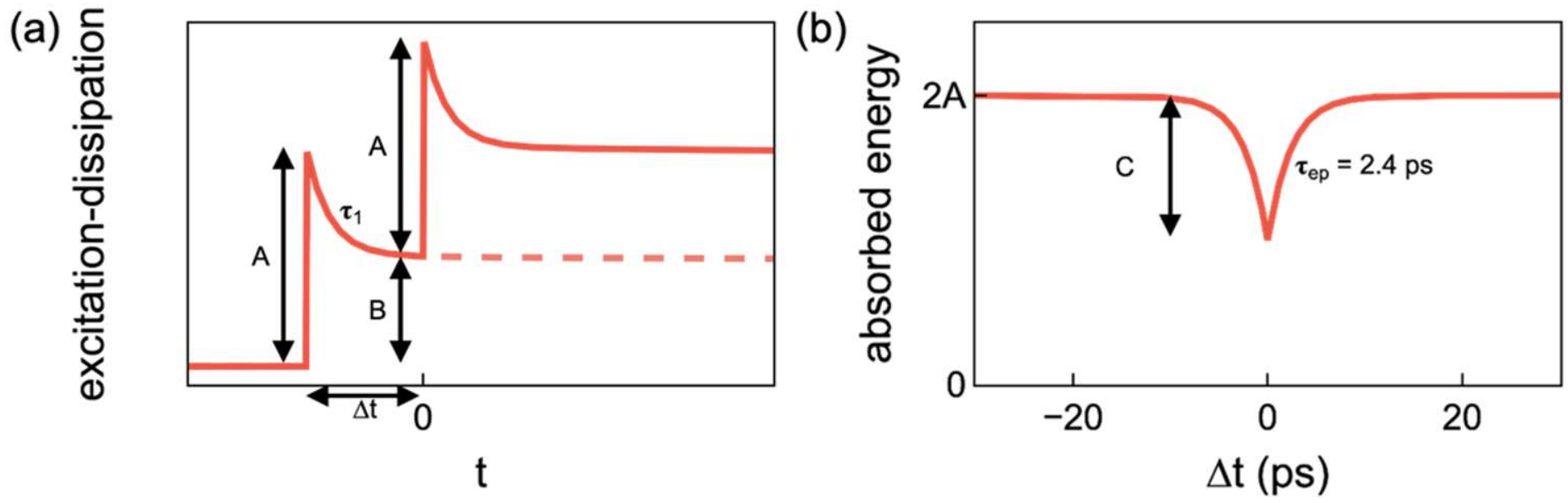

**Fig. S9 | Illustration of Heat Dynamics in a CuMnAs Epilayer. a,** The schematic illustration of the heat dynamics within a CuMnAs epilayer following the incidence of a pair of laser pulses separated by a mutual-time-delay $\Delta t$. Each individual laser pulse delivers heat energy of amplitude $A$, which is subsequently dissipated to the substrate with a thermal relaxation time $\tau_1$. Considering that the dissipation of heat from the substrate to the environment occurs over significantly longer timescales, the residual heat background $B$ can be treated as a constant on ps to ns timescales. **b**, Energy delivered to the epilayer by a pair of laser pulses at 1030 nm. The first pulse transiently increases the CuMnAs reflectivity, which is sensed by the second pulse, that results in a transient reduction of the total absorbed energy.

The heat dissipation model given by Eq. (S4.3) allows us to obtain the thermal relaxation times $\tau_1$ from $F_{\mathrm{eff}}(\Delta t)$ data in Fig. S7d and Fig. S8d for 20-nm-thick and 50-nm-thick CuMnAs epilayers, respectively. However, this model fails in describing the data near the zero-time overlap of the pulse pair (see Insets in Fig. 3a and 3b), necessitating an extension of the model in this ultrafast delay time range, which is based on independent pump-probe measurements of transient reflectivity shown in Fig. 3c. As depicted in the inset of Fig. 3c, at 1030 nm, which is the wavelength of laser pulses used for the single-pulse and double-pulse experiments, absorption of a first (pump) laser pulse transiently increases the CuMnAs reflectivity experienced by the second (probe) pulse, which decays on a ~ ps timescale. The mechanism responsible for this transient ultrafast reflectivity increase can be identified by a non-degenerate pump-probe experiment [S1] depicted in Figs. 3c and 3d (see the Methods section for experimental details). The photoexcitation of a metal by an intense femtosecond laser pulse excites the electron distribution out of equilibrium on a time scale much shorter than the electron–phonon interaction time. The resulting non-thermal population of electrons thermalizes rapidly by electron–electron scattering processes. Consequently, a thermalized electron system, which can be described by a Fermi distribution with an elevated electron temperature, is formed within ≈ 100 fs after the impact of the pump pulse. On a picosecond time scale, the excess energy is dissipated from the electron system to the lattice by electron–phonon scattering processes, which leads to an increase in the lattice temperature. Finally, on a longer time scale, heat diffusion dissipates the excess energy and the metal returns to the equilibrium state. Importantly, all the above effects lead to a change in optical properties and, therefore, the corresponding characteristic time constants can be evaluated from the measured optical transient signals [S1]. Experimentally, these time constants can be straightforwardly identified in a non-degenerate pump-probe experiment where a time evolution of differential reflectivity d$R$/$R$ and differential transmission d$T$/$T$ are measured using probe pulses of various wavelengths ($\lambda$) [S1]. The measured reflectivity dynamics can be reproduced by a phenomenological equation [S1]

$$\Delta R/R\,(\delta t,\lambda) = \left[\alpha(\lambda)\left(1-e^{-\delta t/\tau_{ee}}\right)e^{-\delta t/\tau_{ep}} + \beta(\lambda)\left(1-e^{-\delta t/\tau_{ep}}\right)\right]e^{-\delta t/\tau_1}. \qquad \text{(S4.4)}$$

The first term, with a spectral weight $\alpha$, represents the electronic response with a rise time described by the electron-electron thermalization time $\tau_{ee}$ and decaying by an energy transfer to the lattice with the characteristic electron-phonon relaxation time $\tau_{ep}$. The second term, with a spectral weight $\beta$, describes the lattice heating, with the same time constant time $\tau_{ep}$, and the thermal relaxation time $\tau_l$ represents the heat diffusion from the excited area. From the data depicted in Figs. 3c and 3d we deduced the values of the electron-phonon relaxation time in CuMnAs $\tau_{\rm ep} \approx 2.4$ ps and the heat diffusion time constants $\tau_l$ for various film thicknesses (see

Inset in Fig. 3d), which are fully consistent with the values deduced from the double-pulse experiments.

To be applicable also on ultrafast (ps) times scales, the heat dissipation model given by Eq. (S4.3) has to be modified to include also this transient increase of the sample´s reflectivity, which is reducing the energy absorbed in CuMnAs (with an amplitude $C$ and time constant $\tau_{ep}$) if the delay time between the laser pulse pair is in a ~ ps timescale (see Fig. S9b). Moreover, in the measured double-pulse data (Figs. 3a and 3b) we also identified a presence of an additional component with an amplitude $D$ and time constant $\tau_x$. Overall, the revised heat dynamics model is expressed as:

$$F_{\text{eff}}(\Delta t) \sim \left((A-B)e^{-\frac{\Delta t}{\tau_1}}+B\right)+\left(A-Ce^{-\frac{\Delta t}{\tau_{ep}}}\right)+\left(De^{-\frac{\Delta t}{\tau_x}}\right) \qquad \text{(S4.5)}$$

The fits by this model to the $F_{\text{eff}}(\Delta t)$ data for the 20- and 50-nm-thick CuMnAs epilayers are depicted in Figs. S7d and S8d, respectively. The obtained parameters are summarized in Table S1. At present, we do not have any clear physical interpretation of the origin of the heat dissipation component with the amplitude $D$, which is observed experimentally in the electrical double-pulse experiments on prepared devices but not in the optical pump-probe experiments on the bare epilayers. Further studies are needed to address this.

**Table S1 | Parameters of the heat dissipation model for 20- and 50-nm-thick CuMnAs epilayers.**

| Component | 20 nm | | 50 nm | |
|---|---|---|---|---|
| | Amplitude (mJ/cm$^2$) | Time constant | Amplitude (mJ/cm$^2$) | Time constant |
| A | 18.4 | $\tau_1$ = 375 ps | 18.8 | $\tau_1$ = 1250 ps |
| B | 3.2 | - | 9.0 | - |
| C | 1.7 | $\tau_{ep}$ = 2.4 ps | 1.4 | $\tau_{ep}$ = 2.4 ps |
| D | 2.0 | $\tau_x$ = 20 ps | 0.7 | $\tau_x$ = 20 ps |

**Supplemental references**